\documentclass{sig-alternate}

\usepackage[utf8]{inputenc}
\usepackage{array}
\usepackage[hyphens]{url}
\usepackage{multirow}
\usepackage{tabularx}
\usepackage[tight]{subfigure}
\usepackage{amsmath}
\usepackage{longtable}
\usepackage{dblfloatfix}
\usepackage{balance}

\begin{document}

\title{Semantic Stability in Social Tagging Streams} 

\numberofauthors{2}
\author{
   \alignauthor Claudia Wagner\\
     \affaddr{U. of Koblenz \& GESIS}\\
     \email{claudia.wagner@gesis.org}\\
     \and
   \alignauthor Philipp Singer\\
     \affaddr{Graz University of Technology}\\
     \email{philipp.singer@tugraz.at}\\
   \and
   \alignauthor Markus Strohmaier\\
     \affaddr{U. of Koblenz \& GESIS}\\
     \email{strohmaier@uni-koblenz.de}\\
    \and
   \alignauthor Bernardo Huberman\\
     \affaddr{HP labs Palo Alto}\\
     \email{bernardo.huberman@hp.com}
}
\maketitle

\begin{abstract}
One potential disadvantage of social tagging systems is that due to the lack of
a centralized vocabulary, a crowd of users may never manage to reach a consensus on the
description of resources (e.g., books, users or songs) on the Web.
Yet, previous research has provided interesting evidence that the tag
distributions of resources may become semantically stable over time as more and more users tag them.
At the same time, previous work has raised an array of new questions such as:
(i) How can we assess the semantic stability of social tagging systems in a robust and
methodical way? (ii) Does semantic stabilization of tags vary across
different social tagging systems and ultimately, (iii) what are the factors that
can explain semantic stabilization in such systems?
In this work we tackle these questions by (i) presenting a novel and robust
method which overcomes a number of limitations in existing methods, (ii)
empirically investigating semantic stabilization processes in a wide range of
social tagging systems with distinct domains and properties and (iii) detecting
potential causes for semantic stabilization, specifically imitation behavior,
shared background knowledge and intrinsic properties of natural language.
Our results show that tagging streams which are generated by a \emph{combination of} imitation dynamics and shared background knowledge exhibit faster and higher semantic stability than tagging streams which are generated via imitation dynamics or natural language streams alone.

\end{abstract}






\section{Introduction}
\label{sec:intro}








Instead of enforcing rigid taxonomies or ontologies with
controlled vocabulary, social tagging systems allow users to freely choose
so-called tags to annotate resources on the Web such as users, books or videos.
A potential disadvantage of tagging systems is that due to the lack of a
controlled vocabulary, a crowd of users may never manage to reach a consensus or
may never produce a semantically stable description of resources. By  \emph{semantically stable} we mean that users have agreed on a set of descriptors and their relative importance for a resource which both remain stable over time.
Note, if all descriptors are equally important, users have not produced a shared
and agreed-upon description of a resource, but disagreement may lead to this situation where all descriptors have equally low importance.

Yet, when we observe real-world social tagging processes, we can identify
interesting dynamics from which a semantically stable set of descriptors may emerge for a given
resource.
This semantic stability has important implications for the collective usefulness of
individual tagging behavior since it suggests that information organization
systems can achieve meaningful resource descriptions and interoperability across
distributed systems in a decentralized manner \cite{Macgregor04}.
Semantically stable tagging streams of resources are not only essential for attaining
meaningful resource interoperability across distributed systems and search, but also for learning lightweight semantic models and ontologies from tagging data (see e.g., \cite{Schmitz06miningassociation,Specia2007,Mika2007}) since ontologies are  agreed-upon and shared conceptualizations \cite{Gruber1995}.
Therefore, semantic stability of social tagging streams\footnote{We define a (social) tagging stream as a a temporally ordered sequence of tags that annotate a resource.} is a prerequisite for learning ontologies from tagging data, since it measures the extent to which users have produced a stable and agreed-upon description of a resource.

\begin{figure}[t!]
 \centering
    \vspace{-15pt}
   \begin{tabular}{cc}
\subfigure[Nathan Fillion]{\includegraphics[width=0.24\textwidth]{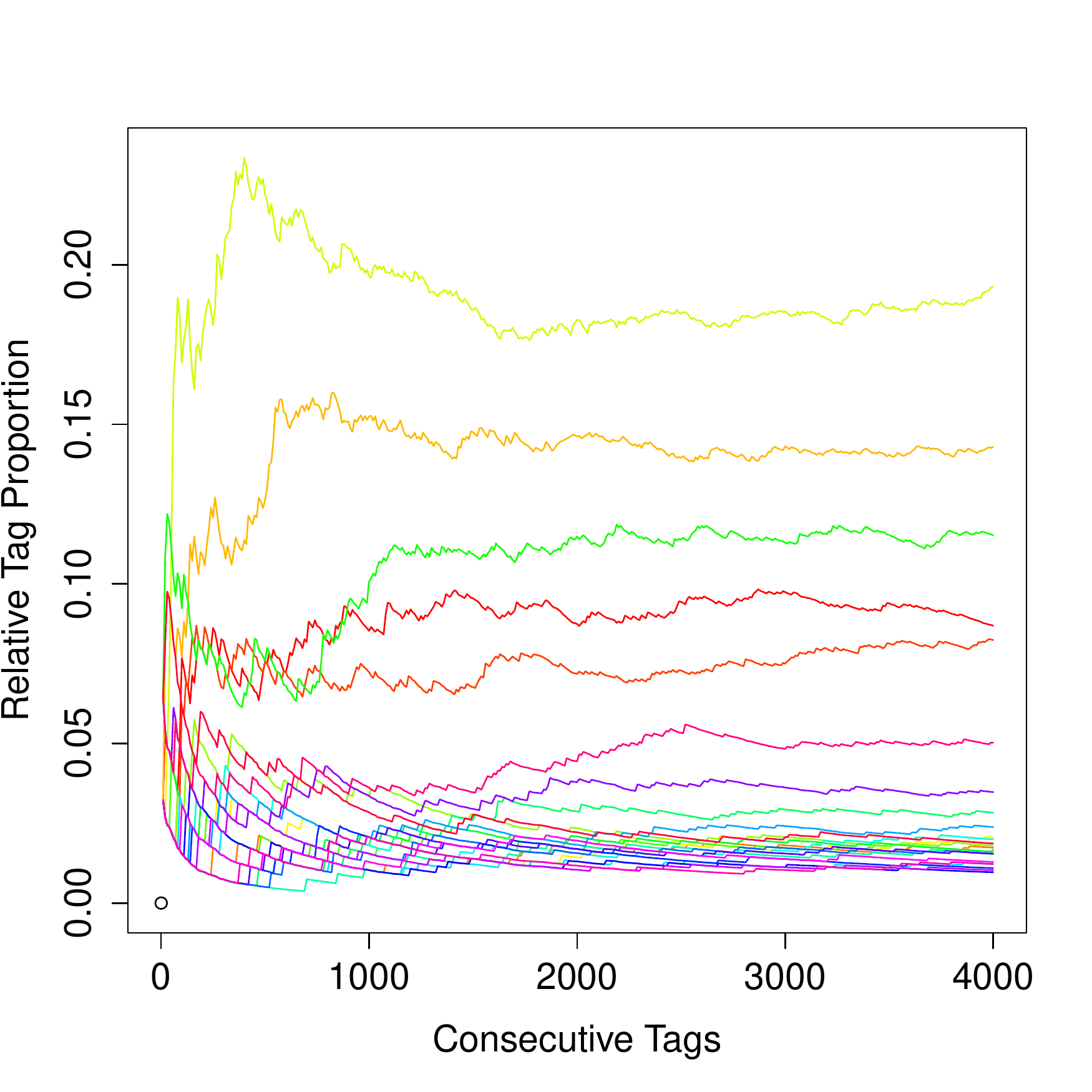}
  \label{NathanFillion_prop_user}}
\subfigure[Sky Sports]{\includegraphics[width=0.24\textwidth]{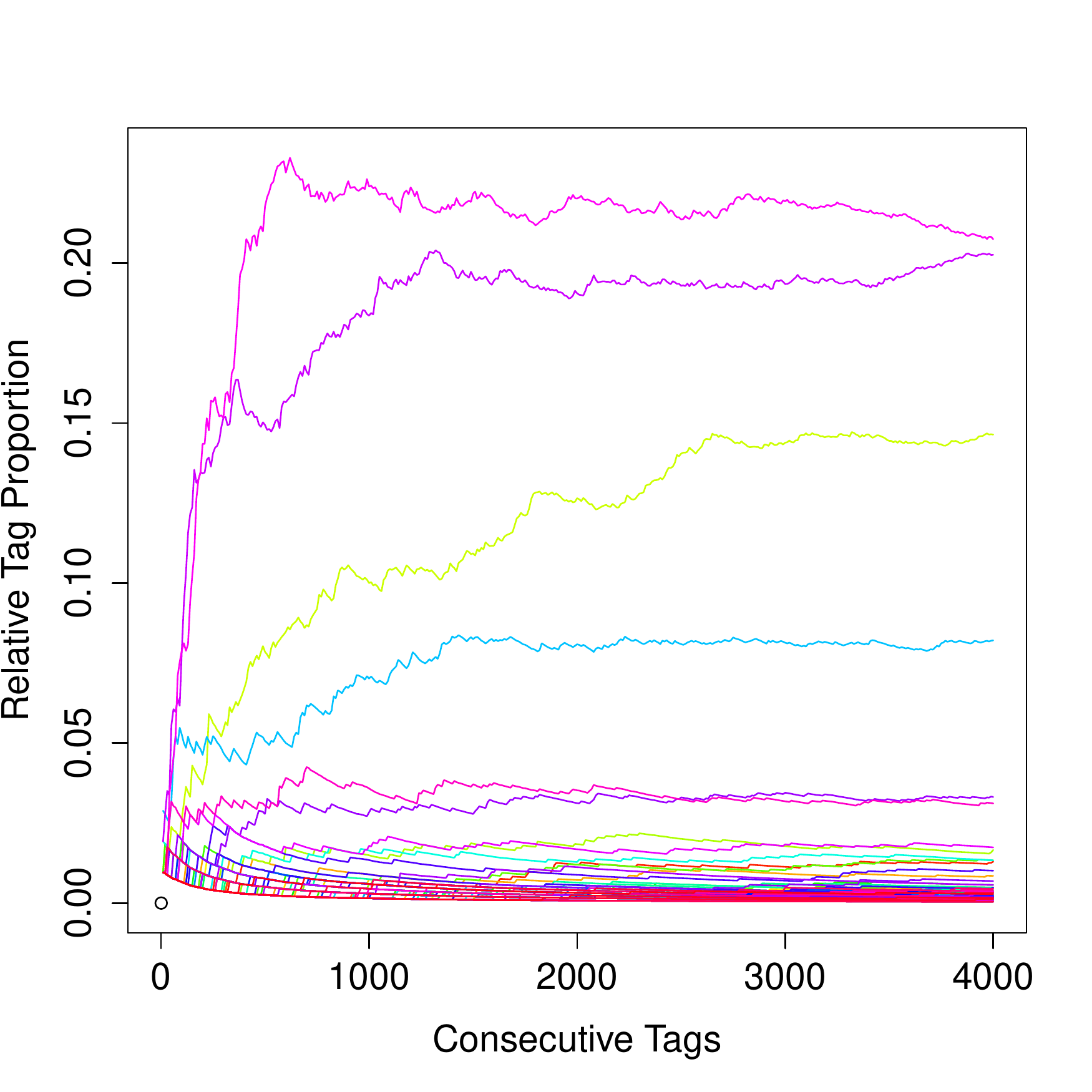}
  \label{SkySports_user_prop_user}}
      \end{tabular}
  \caption{Relative tag proportions of one heavily tagged Twitter user and one
  moderately tagged Twitter user. One can see that the relative tag proportions
  become stable as more users assign tags to the two sample users.}
    \vspace{-10pt}
  \label{fig:props}
\end{figure}

These observations have sparked a series of research efforts focused on (i) methods
for assessing semantic stability in tagging streams (see e.g., \cite{Golder2006,Halpin2007}), (ii) empirical investigations into the semantic
stabilization process and the cognitive processes behind
tagging (see e.g., \cite{Fu2010,Lorince2013}) and (iii) models for
simulating the tagging process (see e.g., \cite{cattuto2007,Dellschaft2008}).
Figure \ref{fig:props} gives an example of such a real world tagging stream and
a corresponding approach to assert semantic stabilization \cite{Golder2006}.
This previous work has proposed to visually analyze the relative tag proportions of
the resource being tagged as a function of consecutive tag assignments. We can
assume that the tagging stream of a resource becomes stable if the relative
tag proportions stop changing.

\noindent {\bf Research questions.}
While previous work makes a promising case for the existence of
semantic stabilization in tagging streams, it raises more questions that require
further attention, including but not limited to the following:
(i) What exactly is semantic stabilization in the context of social tagging
streams, and how can we assert it in a robust way? (ii) How suitable are the different methods which have
been proposed so far and how do they differ? (iii) Does semantic stabilization of resources
vary across different social tagging systems and if yes, in what ways? And finally,
(iv) what are the factors that may explain the emergence of semantic stability
in social tagging streams?


\noindent {\bf Contributions.}
The main contributions of this work are threefold.
We start by making a \emph{methodological} contribution. Based on a systematic
discussion of existing methods for asserting semantic stability in social
tagging systems we identify potentials and limitations.
We illustrate these on a previously unexplored people tagging dataset and a
synthetic random tagging dataset. We explore different subsamples of our dataset including heavily or moderately tagged resources (i.e., a high or moderate amount of
users have tagged a resource).
Using these insights, we present a novel
and flexible method which allows to measure and compare the semantic
stabilization of different tagging systems in a robust way. Flexibility is
achieved through the provision of two meaningful parameters, robustness is
demonstrated by applying it to random control processes.

Our second contribution is \emph{empirical}. We conduct large-scale, empirical
analyses of semantic stabilization in a series of distinct social tagging
systems using our method. We find that semantic stabilization of tags varies
across different systems, which requires deeper explanations of the dynamic
underlying stabilization processes in social tagging systems.

Our final contribution is \emph{explanatory}. We investigate factors
which may explain stabilization processes in social tagging systems. Our results show that tagging streams which are generated by a \emph{combination of} imitation dynamics and shared background knowledge exhibit faster and higher semantic stability than tagging streams which are generated via imitation dynamics or natural language streams alone.


\noindent {\bf Structure.}
This paper is structured as follows: 
We start by discussing related work and methods for measuring semantic
stabilization in tagging systems in Section~\ref{sec:relWork}.  
In Section~\ref{sec:semstab} we highlight that not all state-of-the-art methods are equally suited for measuring semantic stability of
tagging systems, and that some important limitations hinder progress towards a deeper understanding about social-semantic dynamics involved.
Based on this
discussion, we introduce the
data used for our empirical study in Section~\ref{sec:dataset} and present a
novel method for assessing semantic stability and for exploring the semantic stabilization process in Section~\ref{rankMeasure}.
In Section \ref{sec:exsemstab} we aim to shed some light on the
factors which may influence the stabilization process. We discuss and conclude our work in Section~\ref{sec:discuss} and \ref{sec:conclusions}.


\section{Related Work}
\label{sec:relWork}




Social tagging systems have emerged as an alternative to traditional forms of
organizing information which usually enforce rigid taxonomies or ontologies with controlled
vocabulary. Social tagging systems, however, allow users to freely choose
so-called tags to annotate resources such as websites, users, books, videos or artists.

In past research, it has been suggested that stable patterns may emerge when a
large group of users annotates resources on the Web.
That means, users seem to reach a consensus about the description of a resource
over time, despite the lack of a centralized vocabulary which is a central
element of traditional forms of organizing information
\cite{Golder2006,Halpin2007,cattuto2007}. Several methods have been established
to measure this semantic stability: (i) in previous work one co-author of this paper
suggested to assess semantic stability by analyzing the proportions of tags for a given resource as a
function of the number of tag assignments  \cite{Golder2006}.
(ii) Halpin et al.~\cite{Halpin2007} proposed a direct method for quantifying
stabilization by using the Kullback-Leibler (KL) divergence between the rank-ordered tag frequency distributions of a resource at
different points in time. (iii) Cattuto et al.~\cite{cattuto2007} showed that
power law distributions emerge when looking at rank-ordered tag frequency distributions of
a resource which is an indicator of semantic stabilization.


Several attempts and hypotheses which aim to explain the observed stability have emerged. In  \cite{Golder2006} the authors propose that the
simplest model that results in a power law distribution of tags would be the
classic Polya Urn model.
The first model that formalized the notion of new tags was proposed by Cattuto
et al.~\cite{cattuto2007} by utilizing the Yule-Simon
model \cite{Yule1925}. Further models like the semantic imitation model
\cite{Fu2010} or simple imitation mechanisms \cite{Lorince2013} have
been deployed for explaining and reconstructing real world semantic
stabilization.

While above models mainly focus on the imitation behavior of users for
explaining the stabilization process, shared background knowledge might also be
a major factor as one co-author of this work already hypothesized in previous
work \cite{Golder2006}.
Research by Dellschaft et al.~\cite{Dellschaft2008} picked up this hypothesis
and explored the utility of background knowledge as an additional explanatory
factor which may help to simulate the tagging process.
Dellschaft et al. show that combining background knowledge with imitation
mechanisms improves the simulation results. Although their results are very
strong, their evaluation has certain limitations since they focus on reproducing
the sharp drop of the rank-ordered tag frequency distribution between rank 7 and
10  which was previously interpreted as one of the main characteristics of
tagging data \cite{cattuto2006}.
However, recent work by Bollen et al.~\cite{Bollen2009} questions that the
flatten head of these distributions is a characteristic which can be attributed
to the tagging process itself. Instead, it may only be an artifact of the user
interface which suggests up to ten tags.
Bollen et al. show that power law forms regardless
of whether tag suggestions are provided to the user or not, making a strong
point towards the utility of background knowledge for explaining the
stabilization.

In addition to imitation
and background knowledge, an alternative and completely different
explanation for the stable patterns which one can observe in tagging systems
exists, namely the regularities and stability of natural language systems.
Tagging systems are build on top of natural language and if all natural language systems
stabilize over time, also tagging streams will stabilize.
Zipf's law \cite{Zipf1949} states that the frequency of a word in a corpus is
proportional to the inverse of its frequency rank and was found in many different natural language corpora (cf. \cite{montemurro2002})
However, some researcher claim that Zipf's law is inevitable and also a randomly
generated letter sequence exhibits Zipf's law \cite{Chomsky63,Li92randomtexts}.
Recent analysis refuted this claim \cite{Cohen1997,Cancho2010} and
further showed that language networks (based on word co-occurrences) exhibit
small world effects and scale-free degree distributions
\cite{cancho01thesmall}.

While previous work mainly neglected the impact of individual's tagging pragmatics, our previous work showed that not all users contribute equally to the emergence of
tag semantics and that ``describer'' are more useful than ``categorizer'' \cite{Korner2010}. 
Similar to our work \cite{Lin2012} also investigates tag distribution on a macro level (i.e., per system) and on a micro level (i.e., per resource). However, unlike in our work they use distribution fitting to explore the stabilization process. 

\section{State-of-the-art Methods for\\ assessing Semantic Stabilization}
\label{sec:semstab}

In the following, we compare and discuss three existing and well-known
state-of-the-art methods for measuring stability of tag distributions:
\textit{Stable Tag Proportions} \cite{Golder2006}, \textit{Stable Tag
Distributions} \cite{Halpin2007} and \textit{Power Law Fits} \cite{cattuto2007}.
We define the tag distributions as rank-ordered tag frequencies where the
frequency of a tag depends on how many users have assigned the tag to a
resource.
We illustrate the usefulness and limitations of these methods on a previously
unexplored people tagging dataset\footnote{The limitations of the methods are
independent of the dataset and we get similar results using the other datasets
introduced in Section~\ref{sec:dataset}.} and a synthetic random tagging dataset
which will both be described in Section~\ref{sec:dataset}.
Each section (i) points out the intuition and definition of the method, (ii)
applies the method to the data, and (iii) describes limitations and potentials
of the method at hand.

\subsection{Method 1: Stable Tag Proportions \cite{Golder2006}}

\textbf{Intuition and Definition:}
In previous work,  Golder and Huberman \cite{Golder2006} analyzed the relative
proportion of tags assigned to a given resource (i.e., $P(t|e)$ where $t$ is a tag and $e$ is an resource) as a function of the number of tag assignments.
In their empirical study on Delicious the authors found a stable pattern in
which the proportions of tags are nearly fixed after few hundred tag
assignments of each website.

\textbf{Demonstration:}
In Figure~\ref{fig:props} we observe that the tags of different types of
resources (Twitter users rather than websites) also give rise to a stable pattern in which the relative
proportions of tags are nearly fixed.
This indicates that although users keep creating new tags and assign them to
resources, the proportions of the tags per resource become
stable.

  \begin{figure}[t!]
 \centering
 \vspace{-15pt}
\includegraphics[width=0.6\columnwidth]{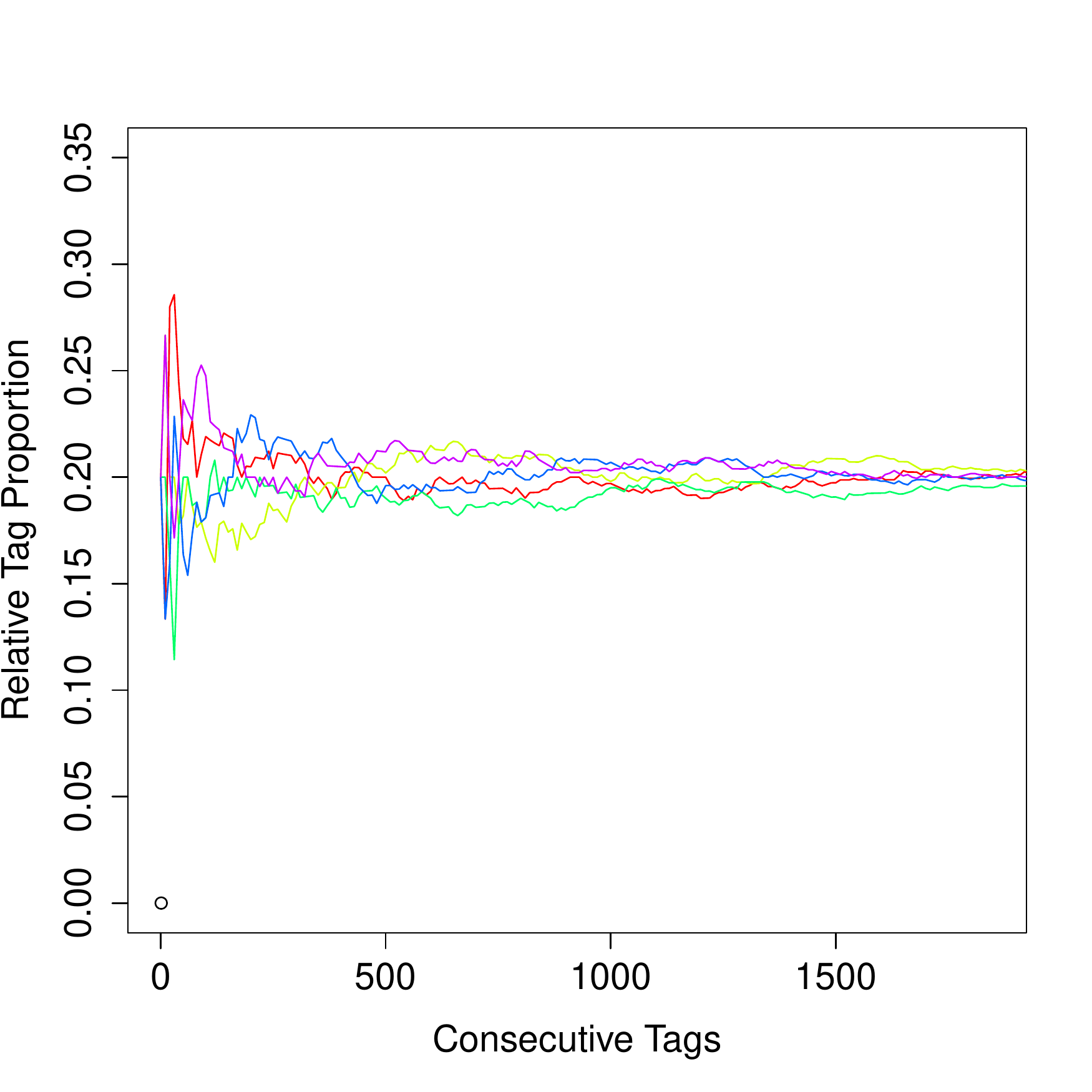}
\vspace{-10pt}
    \caption{Relative tag proportion of a random tagging process where each tag
    assignment on the x-axis corresponds to picking one of the five tags
    uniformly at random. One can see that all tag proportions become relatively stable over time but are all similar.}
    \vspace{-10pt}
  \label{random_prop}
\end{figure}

\textbf{Limitations and Potentials:}
In \cite{Golder2006} the authors suggest that the stability of tag proportions
indicates that users have agreed on a certain vocabulary which describes the resource.
However, also tag distributions produced
by a random tagging process (see Figure \ref{random_prop}) become stable as more
tag assignments take place since the impact of a constant number of tag assignments decreases over time because the total sum of the tag frequency vector increases.

However, the stable tagging patterns shown in Figure~\ref{fig:props}
go beyond what can be explained by a random tagging model, since a random
tagging model produces similar proportions for all tags (see Figure \ref{random_prop}). 
Hence, small changes in the tag frequency vector are enough
to change the order of the ranked tags (i.e., the relative importance of the tags for
the resource).
For real tag distributions this is not the case since these tag distributions
are distributions with short heads and heavy tails -- i.e., few tags
are used far more often than most others.
We exploit this observation for defining our novel method for assessing semantic stability in Section
\ref{rankMeasure}.


\subsection{Method 2: Stable Tag Distributions \cite{Halpin2007}}

\textbf{Intuition and Definition:}
Halpin et al. \cite{Halpin2007} present a method for measuring the semantic
stabilization by using the Kullback Leibler (KL) divergence between the tag
distributions of a resource at different points in time.
The KL divergence between two probability distributions $Q$ and $P$ is defined as follows:
\begin{equation}
D_{KL}(P||Q)= \sum\limits_{i} P(x) ln(\frac{P(x)}{Q(x)})
\end{equation}

The authors use the rank-ordered tag frequencies of
the 25 highest ranked unique tags per resource at different points in time to compute the KL divergence.
The authors use one month as a time window rather than using a fixed number of tag
assignments as Golder and Huberman~\cite{Golder2006} did or we do. This is
important since their measure, per definition, converge towards zero if the number of tag assignments is constant as we will show later.

\begin{table*}[h!b!]
\centering
\vspace{-10pt}
\caption{Parameters of the best power law fits}
\begin{tabular}{|c||c|c||c|c||c|c|} \hline
& $\alpha$ & std & $xmin$ & std & $D$ & std \\ \hline
Heavily tagged users & $1.9793$ &  $0.0841$ & $4.5500$ & $1.9818$ & $0.0299$ &
$0.0118$ \\ \hline
Moderately tagged users & $2.0558$ & $0.1529$ & $3.1200$ & $0.0570$ &
$0.0570$ & $0.0218$ \\ \hline
\end{tabular}
\label{tab:powerlawparam}
\end{table*}

\textbf{Demonstration:}
We use the rank-ordered tag frequencies of the 25 highest ranked tags of each
resource and a constant number ($M$) of consecutive tag assignments. We compare the KL
divergence of tag distributions after $N$ and $N+M$ consecutive tag assignments.
Using a fixed number of consecutive tag assignments allows exploring the
properties of a random tag distribution which is generated by drawing $M$ random
samples from a uniform multinomial distribution.

\begin{figure}[t!]
 \centering
 \begin{tabular}{cc}
  \subfigure[Heavily tagged users]
  {\includegraphics[width=0.24\textwidth]{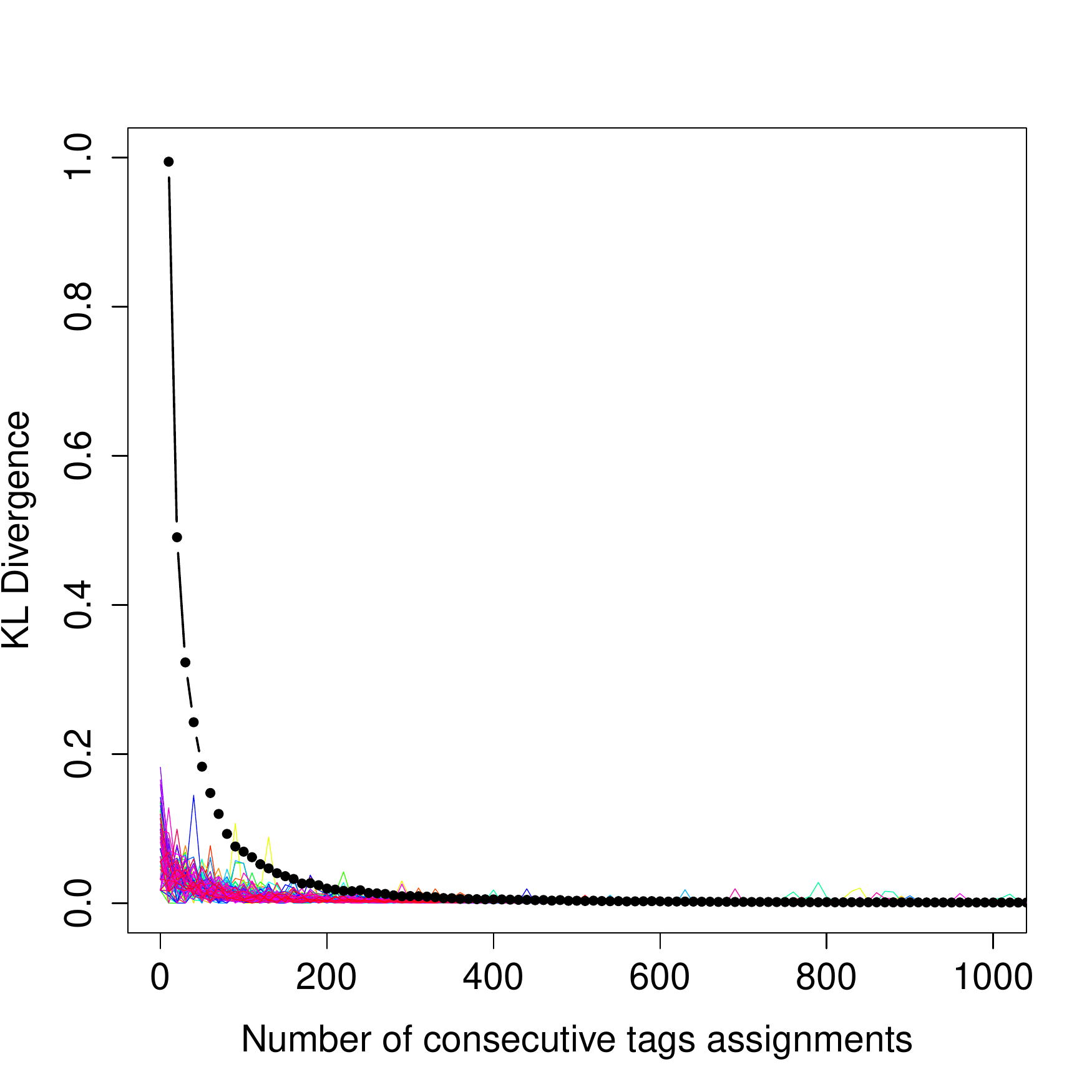}
  \label{twitter_kldiv_user_greater_10k}}
  \subfigure[Moderately tagged users]
  {\includegraphics[width=0.24\textwidth]{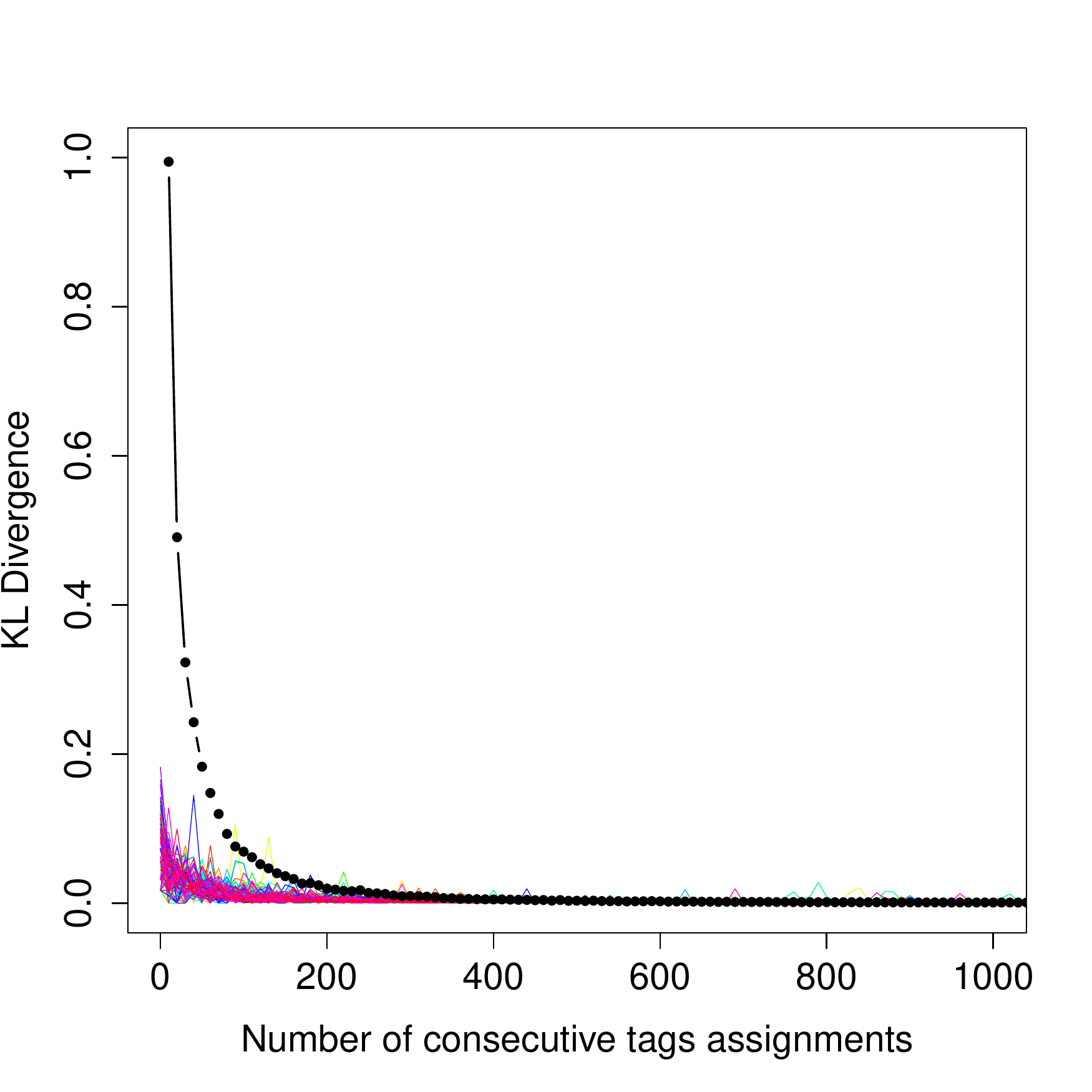}
  \label{twitter_kldiv_user_less_10k}}
\end{tabular}
  \caption{ KL divergence between the tag distributions at consecutive time
  points. Each colored line corresponds to one Twitter user, while the black
  dotted line depicts a randomly simulated tag distributions. One can see
  that the KL divergence decreases as a function of the number of tag assignments.
  The KL divergence of a random tagging process decreases slightly slower than the KL divergence of the real tagging data.}
  \vspace{-10pt}
  \label{fig:kldivs}
\end{figure}

In Figure \ref{fig:kldivs}, each point on the x-axis consists of $M=10$
consecutive tag assignments and $N$ ranges from $0$ to $1000$. The black dotted line indicates the KL divergence of a random tag distribution.
One can see from this figure that not only the tag distributions of resources seem to converge towards zero over time (with few outliers), but also random tag distributions do.

\textbf{Limitations and Potentials:}
A single tag assignment in month $j$ has more impact on
the shape of the tag distribution of a resource than a single tag added in
month $j+1$, if we assume the number of tags which are added per month is
relatively stable over time. However, if the number of tag assignments per resource varies a lot across different
months, convergence can be interpreted as semantic stabilization.

This suggests that without knowing the frequencies of tag assignments per
month, the measure proposed by Halpin et al.~\cite{Halpin2007} is limited with
regard to its usefulness since one never knows whether stabilization can be
observed due to the fact that users agreed on a certain set of descriptors and their relative importance for the resource or due to the fact
that the tagging frequency in later months was lower than in earlier months.
In our work (see Figure~\ref{fig:kldivs}), we compare the KL divergence of a
randomly generated tag distribution with the KL divergence of real tag
distributions. This reveals how much faster users reach consensus compared to what one would expect.

Even though we believe this method already improves the original approach suggested by
Halpin et al.~\cite{Halpin2007}, it is  still limited because it requires to limit the analysis to the top
$k$ tags. The KL divergence is only defined between two distributions over the same set of tags.
We address this limitation with the new method which
we propose in Section \ref{rankMeasure}.


  \begin{figure*}[ht!]
 \centering
 \begin{tabular}{cccc}
   \subfigure[Heavily tagged
  users]{\includegraphics[width=0.24\textwidth]{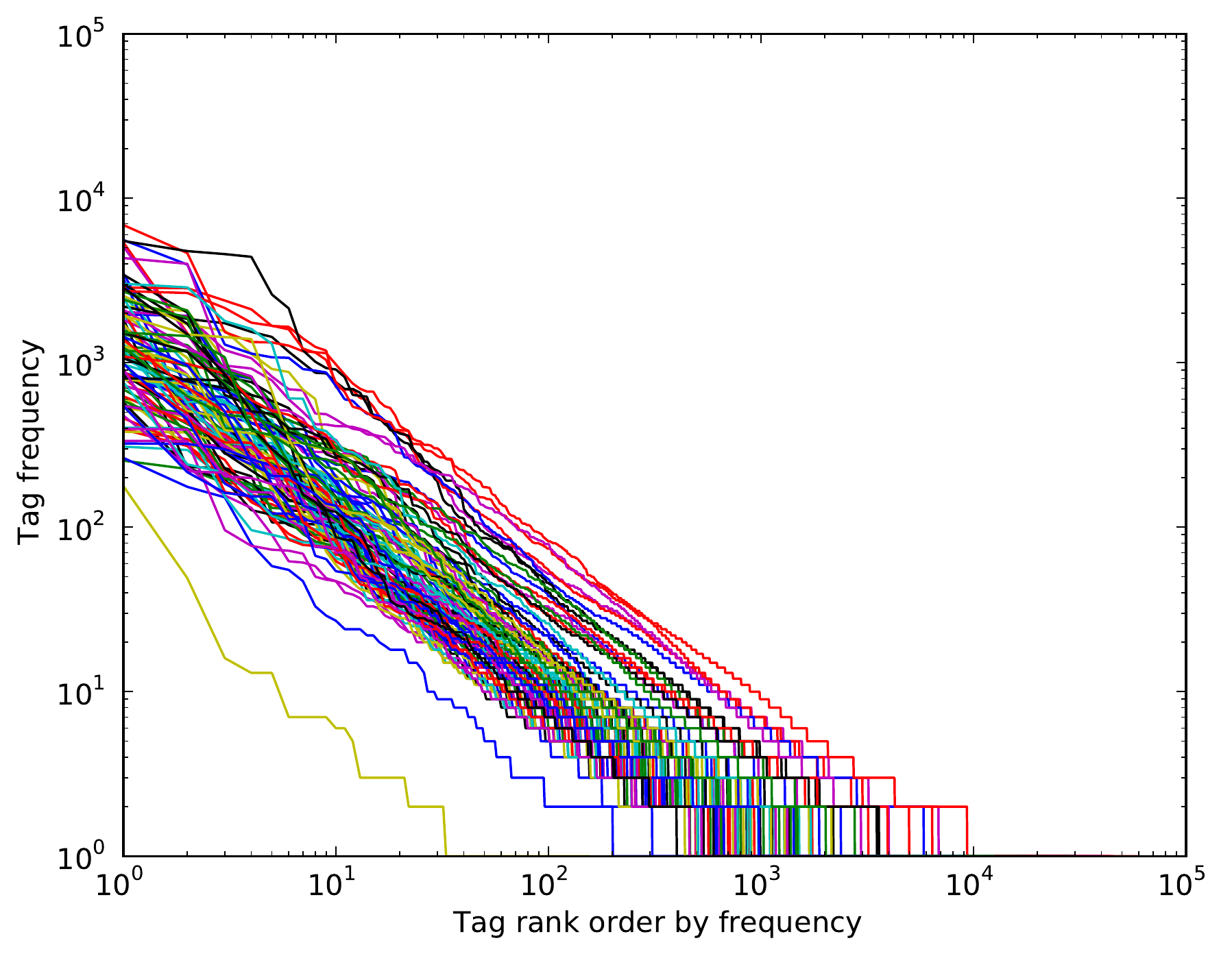}
  \label{subfig:gr10000_tagfrequ}}
  \subfigure[Moderately tagged
  users]{\includegraphics[width=0.24\textwidth]{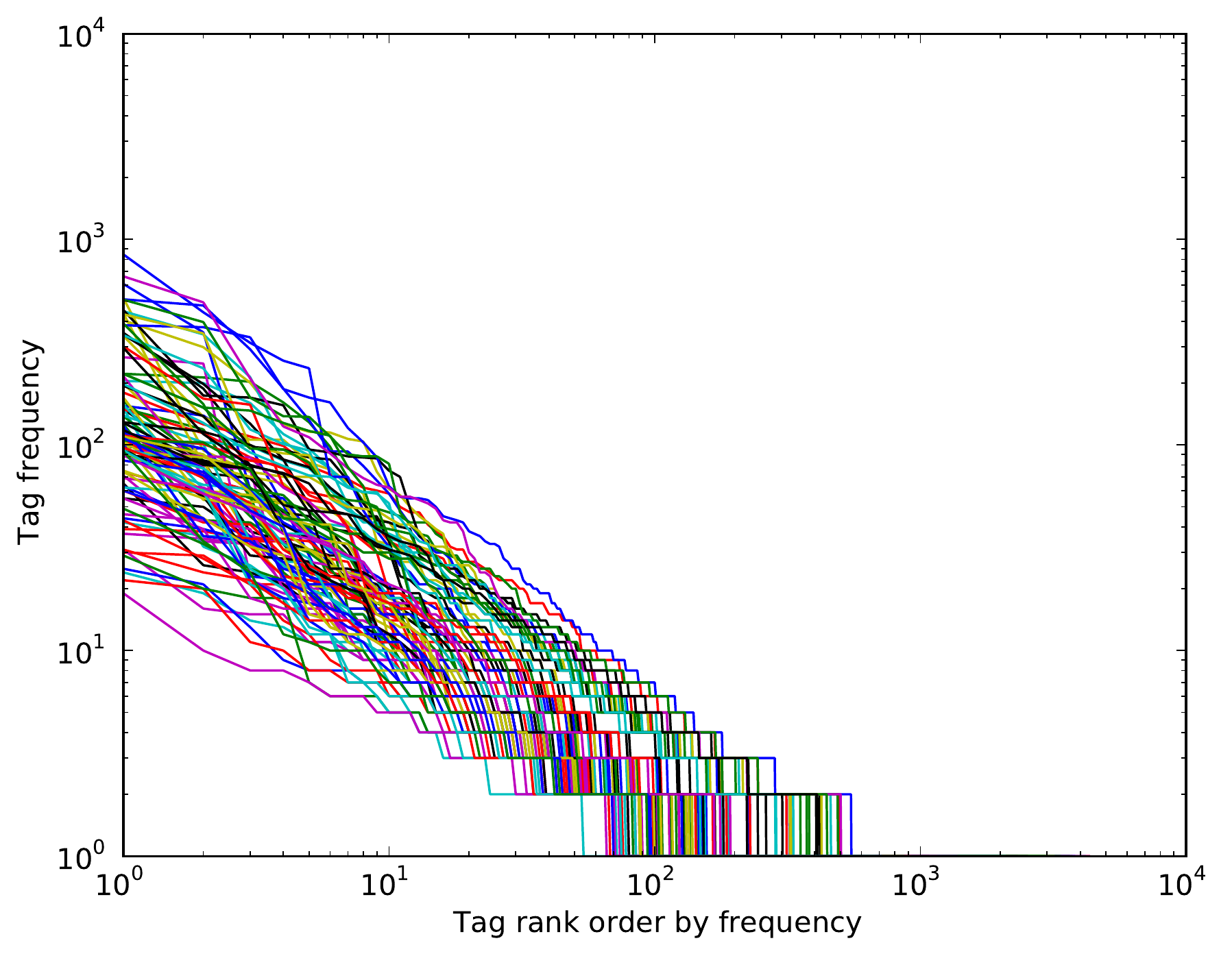}\label{subfig:less10000_tagfrequ}}

  \subfigure[Heavily tagged
  users]{\includegraphics[width=0.24\textwidth]{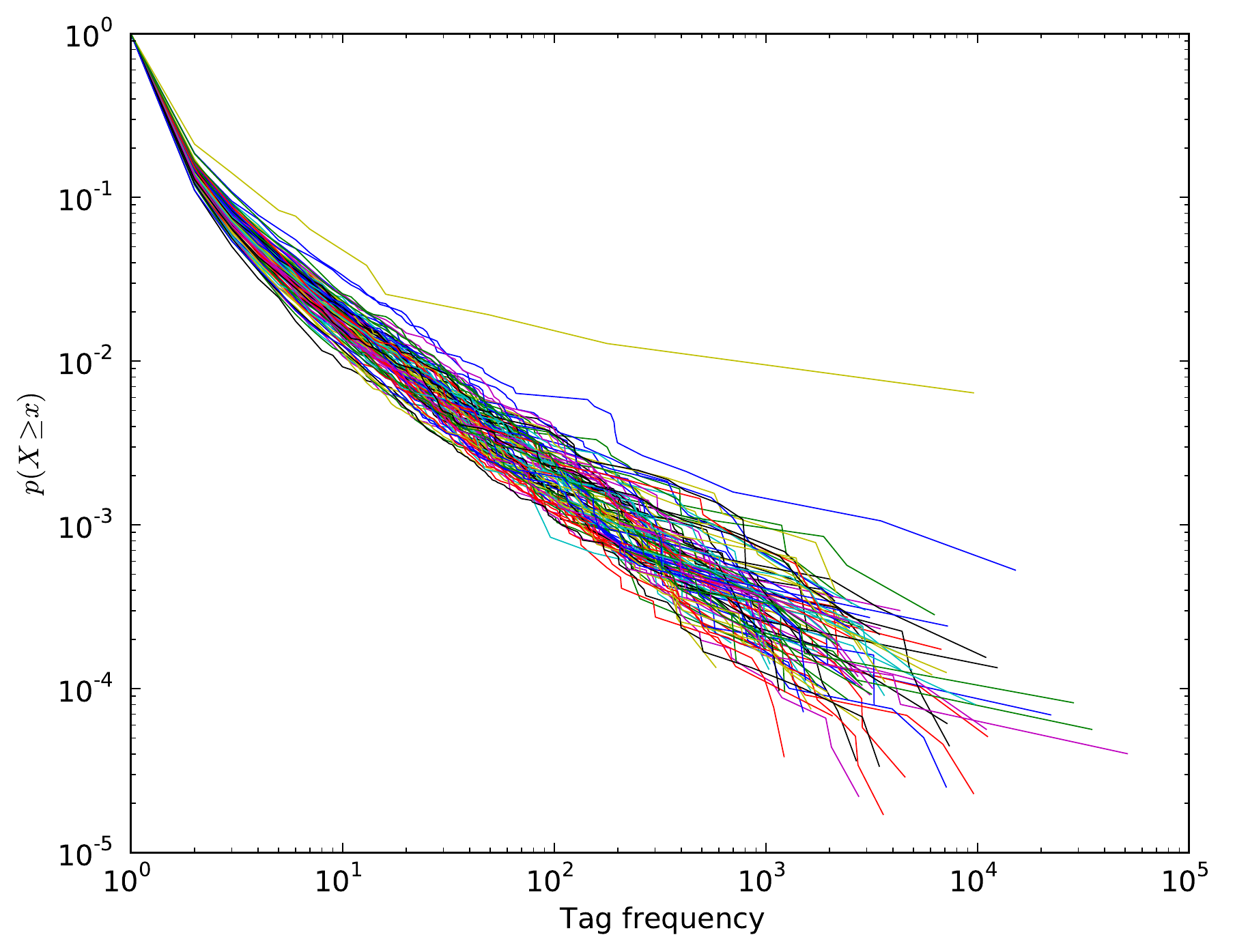}
  \label{subfig:gr10000_ccdf}}
  \subfigure[Moderately tagged
  users]{\includegraphics[width=0.24\textwidth]{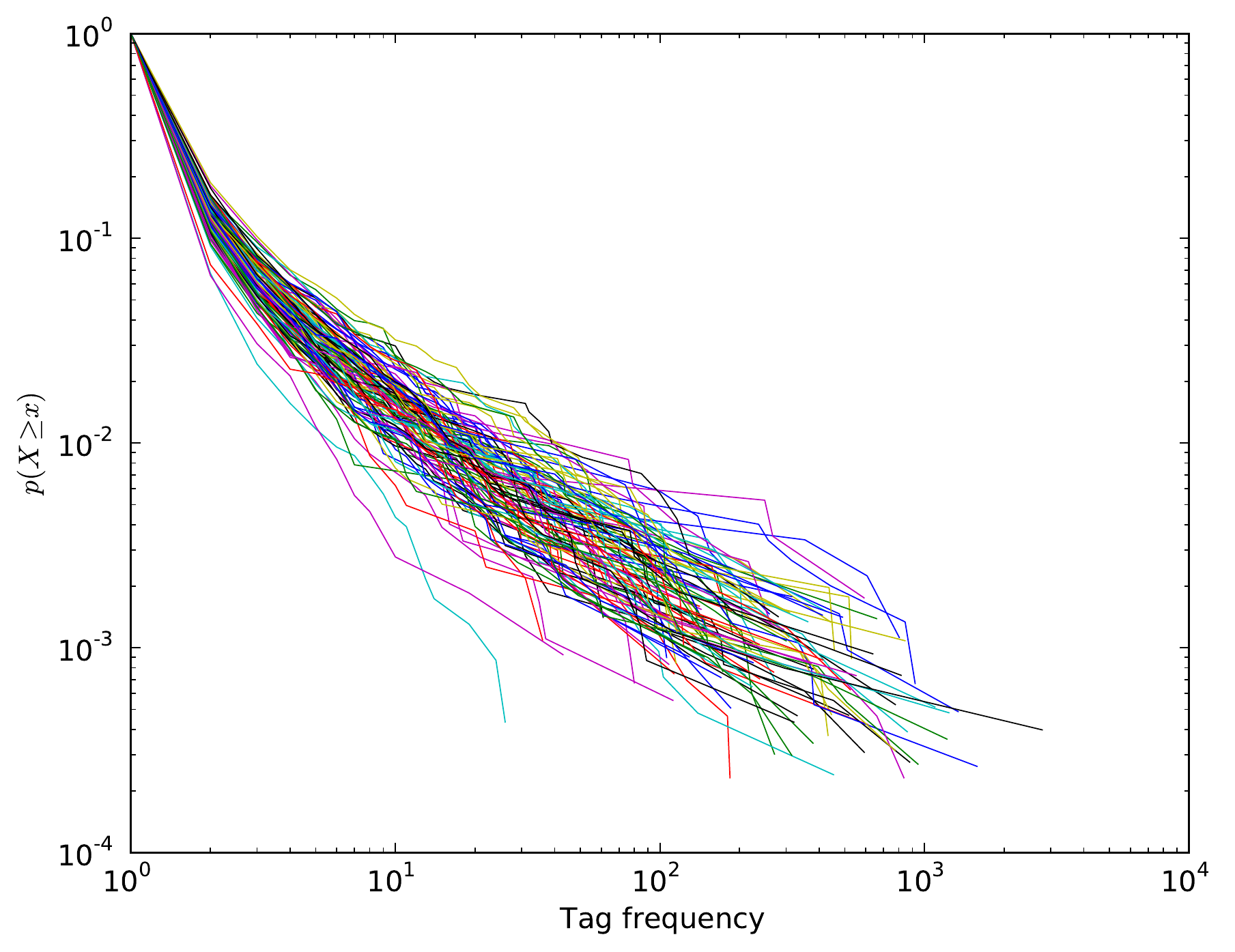}\label{subfig:less10000_ccdf}}

 \end{tabular}
  \caption{Rank-ordered tag frequency and CCDF plots for heavily tagged and
  moderately tagged users on log-log scale. The illustrations show that for both
  heavily and moderately tagged resources, few tags are applied very frequently
  while the vast majority of tags is applied very rarely. In Figure~\ref{subfig:gr10000_ccdf} and
Figure~\ref{subfig:less10000_ccdf} we can see that a large number of tags are
only used once. The figures visualizes that the tails of the tag distributions
are close to a straight line which suggests that the distributions might follow
a power law.}
  \vspace{-10pt}
  \label{fig:ccdf}
\end{figure*}


\subsection{Method 3: Power Law Fits \cite{Mathes2004}}

\textbf{Intuition and Definition:}
Tag distributions which follow a power law are sometimes regarded as semantically stable,
(i) because of the scale invariance property of power law
distributions -- i.e., that regardless how large the system grows, the slope of
the distribution would stay the same, and (ii) because power law distributions
are heavy tail distributions -- i.e., few tags are applied very frequently while
the majority of tags is hardly used.
Adam Mathes~\cite{Mathes2004} originally hypothesized that tag distributions in
social tagging systems follow a power law function. Several studies empirically show that the tag distributions of resources in social tagging systems indeed follow a power law \cite{sen2006,Kipp2006,Cattuto06,cattuto2007}.
A power law distribution is defined by the function:
\begin{equation}
y = cx^{-\alpha} + \epsilon
\end{equation}
Both $c$ and $\alpha$ are constants characterizing the power law
distribution and $\epsilon$ represents the uncertainty in the observed values.
The most important parameter is the scaling parameter $\alpha$ as it represents
the slope of the distribution \cite{Bollen2009,Clauset2009}. It is also
important to remark that real world data nearly never follows a power law for the
whole range of values. Hence, it is necessary to find some minimum value $xmin$ for which one can say that the tail of the
distribution\footnote{we use the term \emph{tail} to characterize the
end of a distribution in the sense of probability theory} with $x  \geq xmin$ follows a power law \cite{Clauset2009}.

\textbf{Demonstration:}
We first visualize the rank frequency tag distributions (see
Figure~\ref{subfig:gr10000_tagfrequ} and Figure~\ref{subfig:less10000_tagfrequ})
and the complementary cumulated distribution function (CCDF) of the probability
tag distributions (see  Figure~\ref{subfig:gr10000_ccdf} and
Figure~\ref{subfig:less10000_ccdf})  on a log-log scale.
We see that for heavily and moderately tagged resources, few
tags are applied very frequently while the vast majority of tags are used very
rarely. Figure~\ref{subfig:gr10000_ccdf} and Figure~\ref{subfig:less10000_ccdf}
show that the tag distributions of heavily and moderately tagged resources are
dominated by a large number of tags which are only used once.

Figure~\ref{fig:ccdf} reveals that the tails of
the tag distributions (starting from a tag frequency $2$)  are
close to a straight line.
The straight line, which is a main characteristic for power law distributions
plotted on a log-log scale, is more visible for heavily tagged resources than for moderately tagged once.
We can now hypothesize that
a power law distribution could be a good fit for our data if we look at the tail
of the distribution with a potential $xmin \geq 2$.

For finding the scaling parameter $\alpha$ we use a \emph{maximum likelihood
estimation} and for finding the appropriate $xmin$ value we use  the \emph{Kolmogorov-Smirnov statistic} as suggested by Clauset et
al. \cite{Clauset2009}.
As proposed in previous work \cite{Bollen2009,Clauset2009}, we also look at the
Kolmogorov-Smirnov distance $D$ of the corresponding fits -- the smaller $D$ the
better the fit.
Table~\ref{tab:powerlawparam} shows the parameters of the best power law fits,
averaged over all heavily tagged or moderately tagged resources. One can see
from this table that the $\alpha$ values are very similar for both datasets and
also fall in the typical range of power law distributions.
Further, one can see that the power law fits
are slightly better for heavily tagged resources than for moderately tagged
once, as also suggested by Figure~\ref{fig:ccdf}.

Although our results suggest that it is likely that our distributions have
been produced by a power law function, further investigations are warranted to explore whether other heavy-tailed
 candidate distributions are better fits than the power law \cite{Clauset2009,Alstott2013}.
We compare our power law fit to the fit of the \emph{exponential
function}, the \emph{lognormal function} and the \emph{stretched exponential
(Weibull) function}.
We use \emph{log-likelihood ratios} to indicate which fit is better.

The exponential function represents the absolute minimal candidate function to
describe a heavy-tailed distribution. That means, if the power law function
is not a better fit than the exponential function, it is difficult to assess whether the
distribution is heavy-tailed at all. The lognormal and stretched exponential function
represent more sensible heavy-tailed functions.
Clauset et al. \cite{Clauset2009} point
out that there are only a few domains where the power law function is a better
fit than the lognormal or the stretched exponential.

Our results confirm this since we do not find significant differences between
the power law fit and the lognormal fit (for both heavily and moderately tagged users).
However, most of the time the power law function is significantly better than
the stretched exponential function and the power law function is a significantly better fit than
the exponential function for all heavily tagged users and for most  moderately tagged users.
This indicates that the tag distributions of heavily tagged resources and most moderately tagged resources are clearly heavy tail distributions and the
power law function is a reasonable well explanation.
However, it remains unclear from which heavy tail distribution the data has been drawn since several of them
produce good fits.


\textbf{Limitations and Potentials:}
As we have shown, one limitation of this method is that it is often difficult to determine which distribution has generated the data since several
distributions with similar characteristics may produce an equally good fit.
Furthermore, the automatic calculation of the best $xmin$
value for the power law fit has certain consequences since $xmin$ might become
very large and therefore the tail to which the power law function is fitted may
become very short.
Finally, there is still an ongoing discussion about the
informativeness of scaling laws (see \cite{Kello2010} for a good overview), since
some previous work suggests that there exist many ways to produce scaling laws and some of those ways are
idiosyncratic and artifactual \cite{Rapoport1982,Li92randomtexts}.

\newpage
\section{Experimental setup and datasets}
\label{sec:dataset}




We conduct large-scale, empirical analyses on the semantic stabilization process in a
series of different social tagging systems using the state-of-the-art methods
described in Section~\ref{sec:semstab} and using a new method introduced in
Section~\ref{rankMeasure}.
Table~\ref{tab:dataset} gives an overview of the datasets obtained from
distinct tagging systems using the nature of the resource being tagged, the
sequential order of the tagging process (i.e., is the resource selected first or the tag), the existence or absence of tag suggestions and the
visibility of the tags  which
have been previously assigned to a resource. We say that tags have a low visibility if users
do not see them during the tagging process and if they are not shown on the page
of the resource being tagged. Otherwise, tags have a high visibility. Further,
the number of resources, users and tags per dataset are shown.

\begin{table*}[h!b!]
\centering
\vspace{-10pt}
\caption{Description of the datasets and characteristics of the social tagging system the data stem from.}
\begin{tabular}{|c|c|c|c|c||c|c|c|} \hline
\textbf{System} & \textbf{Entity Type} &  \textbf{Tag First} & \textbf{Tag Suggestions} & \textbf{Tags Visible} &  \textbf{\#Resources} & \textbf{\#Users} & \textbf{\#Tags} \\ \hline
Delicious & websites &   no & yes &  low  & 17,000k & 532k & 2,400k\\ \hline
LibraryThing & books &  no & no & high  & 3,500k & 150k & 2,000k\\ \hline
Twitter lists & users &   yes & no & low  & 3,286 & 2,290k & 1,111k 
\\ \hline
\end{tabular}
\label{tab:dataset}
\vspace{-0mm}
\end{table*}

\textbf{Delicious dataset:} Delicious is a social tagging system where users can
tag any type of website. We use the Delicious dataset crawled by G\"{o}rlitz et
al.~\cite{Gorlitz2008}.
From this dataset we randomly selected 100 websites which were tagged by many
users (more than 4k users) and 100 websites which were moderately tagged (i.e.,
by less than 4k but more than 1k users) and explore the consecutive tag
assignments for each website. The original dataset is available
online\footnote{\url{http://www.uni-koblenz-landau.de/koblenz/fb4/AGStaab/Research/DataSets/PINTSExperimentsDataSets}}.

\textbf{LibraryThing dataset:} LibraryThing is a social tagging system which
allows to tag books. We use the LibraryThing dataset which was crawled by Zubiaga et
al.~\cite{Zubiaga2011}.
Again, we randomly sampled 100 books that were heavily tagged (more than 2k
users) and 100 books which were moderately tagged (less than 2k and more
than 1k users) and explore the consecutive tag assignments for each book.

\textbf{Twitter dataset:} 
Twitter is a microblogging service that allows users to tag their contacts by
grouping them into user lists with a descriptive title. The creation of such
list titles can be understood as a form of tagging since list titles are free
form words which are associated with one or several resources (in this case
users).
What is unique about this form of tagging is that the tag (aka the list title)
is usually produced first, and then users are added to this list, whereas in
more traditional tagging systems such as Delicious, the process is the other way
around.
From a Twitter dataset which we described in previous work \cite{Wagner2013HP},
we selected a sample of 100 heavily tagged users (which are mentioned in more
than 10k lists) and 100 moderately tagged users (which are mentioned in less
than 10k lists and more than 1k lists). For each of these sample users we
crawled the full history of lists to which a user was assigned. We do not know
the exact time when a user was assigned to a list but we know the relative order
in which a user was assigned to different lists. Therefore, we can study the
tagging process over time by using consecutive list assignments as a sequential
ordering\footnote{We share the Twitter user handles to allow other researchers to recreate our
dataset and reproduce our results for our heavily tagged
\url{http://claudiawagner.info/data/gr_10k_username.csv} and moderately tagged
\url{http://claudiawagner.info/data/less_10k_username.csv} Twitter users.}.


It needs to be noted that the thresholds we have used above during the data
collection are distinct for each tagging system since those systems differ
amongst others in their number of active users and size. We chose the thresholds
empirically and found that the choice of threshold does not impact our
results since heavily tagged as well as moderately tagged resources show
similar characteristics.


Finally, we also contrast our tagging datasets with a natural language corpus
(see Section~\ref{subsec:naturallan}) and a random tagging dataset. This allows
us on one hand, to explore to what extent semantic stabilization which can
be observed in tagging systems goes beyond what one would expect to observe if the
tagging process would be a random process; and on the other hand, to compare the
semantic stabilization of the tag distributions of resources with the semantic
stabilization of co-occurring word distributions of resources.

\textbf{Natural Language corpus:} As a natural language corpus we use a sample
of tweets which refer to the same resource. Therefore, we selected a random
sample of users from our Twitter dataset which have received tweets from many distinct
users (more than 1k). For those users, we select a sample of up to 10k
tweets they received. The words in those tweets are extracted and interpreted as
social annotations of the receiver. This allows us to compare tags with words,
both annotating a resource (in this case a user).

\textbf{Synthetic random tagging dataset:} Given a fixed vocabulary size we can create a random
tagging dataset by simulating the tagging process as random draws from a urn (containing
all possible tags of the vocabulary) where each ball (i.e.,
tag) is returned to the urn after each draw.


\section{Measuring Semantic Stability}
\label{rankMeasure}

Based on the analysis of state-of-the-art methods presented in Section~\ref{sec:semstab}, we (i) present a novel method for assessing the semantic
stability of individual tagging streams and (ii) show how this method can be
used to assess and compare the stabilization process in different tagging systems.
Our new method incorporates three new ideas:

\textbf{Ranking of tags:} A tagging stream can be considered as semantically stable if users have agreed on a ranking of tags which remains stable over time.
It is more important that the ranking of frequent tags
remains stable than the ranking of less frequent tags since frequent tags are
those which might be more relevant for a resource. Frequent tags have been
applied by many users to a resource and therefore stable rankings of these tags
indicate that a large group of users has agreed on the relative importance of
the tags for that resource.

\textbf{Random baselines:} Semantic stability of a random tagging process needs
to be considered as a baseline for stability since we are interested in
exploring stable patterns which go beyond what can be explained by a random tagging process.

\textbf{New tags over time:} New tags can be added over time and therefore, a
method which compares the tag distributions of one resource at different points in
time must be able to handle mutually non-conjoint tag distributions -- i.e.,
distributions which contain tags that turn up in one distribution but not in the
other one. 
Most measures used in previous work (e.g., the KL divergence)
only allow to compare the agreement between mutually conjoint lists of elements
and a common practice is to prune tag distributions to their top k elements --
i.e., the most frequently used tags per resource. However, this pruning requires
global knowledge about the tag usage and only enables a post-hoc rather than a
real-time analysis of semantic stability.

\subsection{Rank Biased Overlap: $RBO(\sigma1, \sigma2, p)$}
\label{subsec:rbo}

\textbf{Intuition and Definition:}
The Rank Biased Overlap (RBO) \cite{webber2010} measures the similarity between
two rankings and is based on the cumulative set overlap. The set overlap at
each rank is weighted by a geometric sequence, providing both top-weightedness
and convergence.
 RBO is defined as follows:
\begin{equation}
RBO(\sigma1, \sigma2, p) = (1-p) \sum_{d=1}^{\infty} \frac{\sigma1_{1:d} \cap \sigma2_{1:d} }{d} p^{(d-1)}
\end{equation}
Let $\sigma1$ and $\sigma2$ be two not necessarily conjoint lists of ranking. Let $\sigma1_{1:d}$ and $\sigma2_{1:d}$ be the ranked lists at depth $d$.
The RBO falls in the range $[0,1]$, where 0 means disjoint, and 1 means identical.
The parameter $p$ ($0 \leq p < 1$) determines how steep the decline in weights
is.
The smaller $p$ is, the more top-weighted the metric is. If $p = 0$, only the
top-ranked item of each list is considered and the RBO score is either zero or one.
On the other hand, as $p$ approaches arbitrarily close to 1, the weights become arbitrarily flat.
These weights, however, are not the same as the weights that the elements at different ranks d themselves take, since these elements
contribute to multiple agreements.

In the following, we use a version of RBO that accounts for tied ranks. As
suggested in \cite{webber2010}, ties are handled by assuming that if $t$ items
are tied for ranks d to $d + (t - 1)$, they all occur at rank $d$. RBO may
account for ties by dividing twice the overlap at depth d by the number of items
which occur at depth d, rather than the depth itself:
\begin{equation}
RBO(\sigma1, \sigma2, p) = (1-p) \sum_{d=1}^{\infty} \frac{2* \sigma1_{1:d} \cap \sigma2_{1:d} }{|\sigma1_{1:d} + \sigma2_{1:d}|} p^{(d-1)}
\end{equation}
We modify RBO by summing only over occurring depths rather than all possible
depths.
Therefore, our RBO measure further penalizes ties and assigns a
lower RBO value to pairs of lists containing ties.
For example, consider the following two pairs of ranked lists of items:
(i) \textit{(A=1, B=2, C=3, D=4)}, \textit{(A=3, B=2, C=1, D=4)} and (ii) \textit{(A=1, B=1, C=1, D=4)}, \textit{(A=1, B=1, C=1, D=4)}.
Both pairs of lists have the same concordant pairs: (A,D) and (B,D) and (C,D).
The RBO value of the first pair is $0.2$ according to the original measure and also according to our tie-corrected variant.
The RBO value of the second pair is $0.34$ according to the original measure and $0.17$ according to our tie-corrected variant.
This example nicely shows that while the original RBO measure tends to
overestimate ties, our variant slightly penalizes ties.
For our use case this makes sense since we do not want to overestimate the
semantic stability of a resource where users have not agreed on a ranking of tags but only find that all of tags are equally important.


  \begin{figure}[t!]
 \centering
  \begin{tabular}{cc}
  \subfigure[Heavily tagged
  users]{\includegraphics[width=0.24\textwidth]{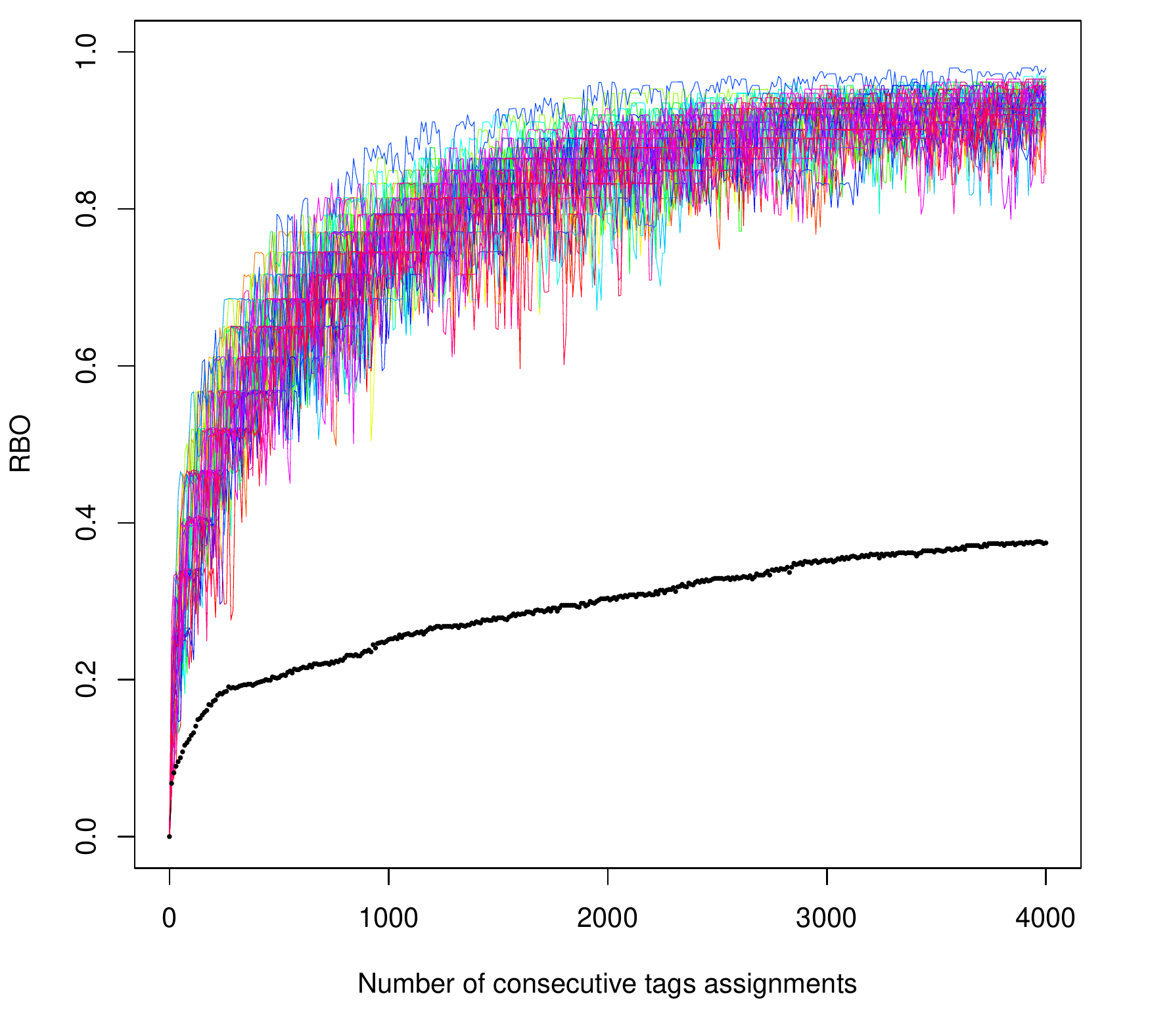}
    \label{subfig:1rbo_twitter}}
  \subfigure[Moderately tagged
  users]{\includegraphics[width=0.24\textwidth]{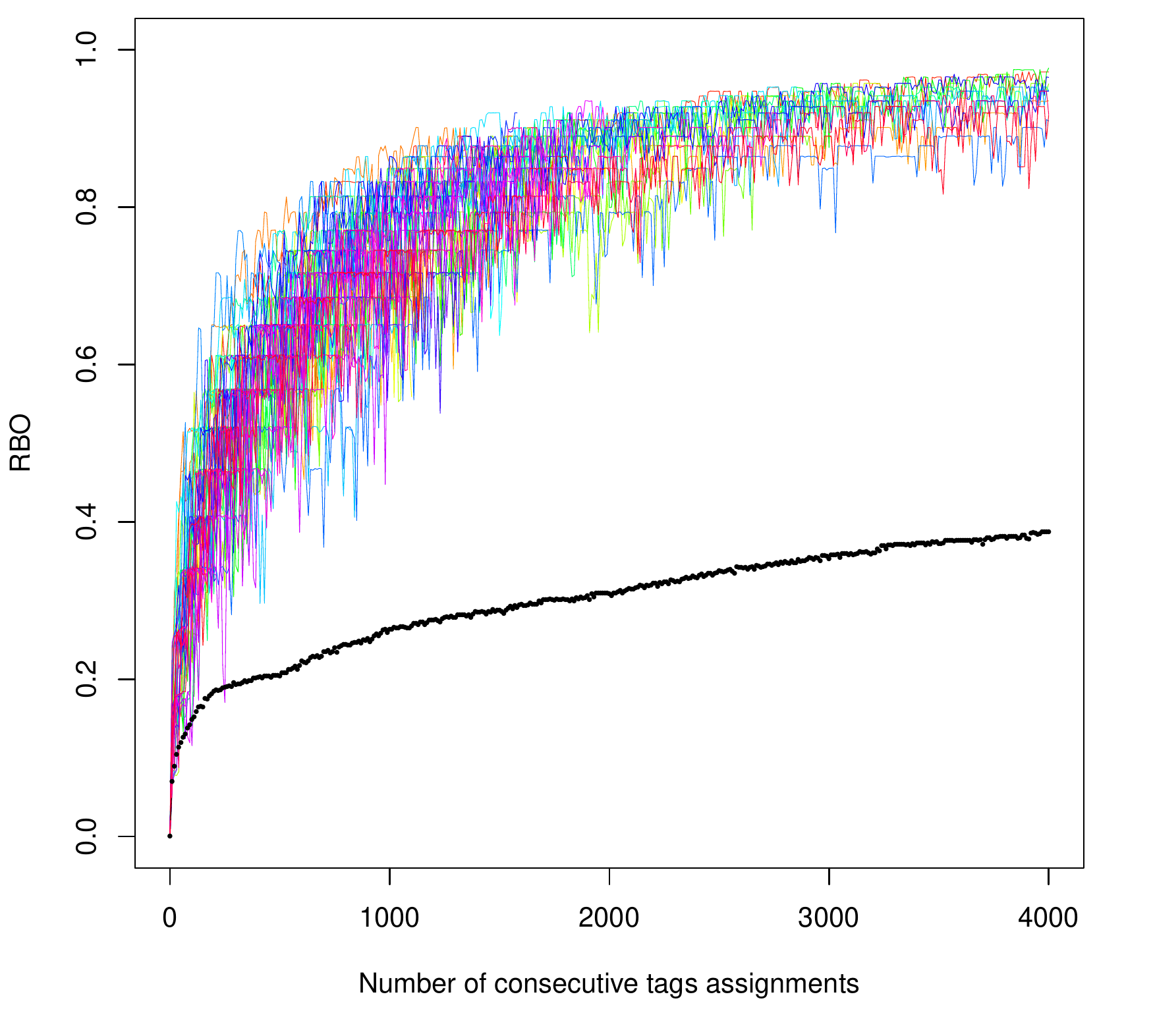}
      \label{subfig:2rbo_twitter}}

   \end{tabular}
  \caption{Rank Biased Overlap (RBO) measures with $p=0.9$. The black dotted
  line shows the weighted average RBO of a random tagging process over time,
  while each colored line corresponds to the RBO of one Twitter user. }
  \label{fig:rbo}
  \vspace{-10pt}
\end{figure}

\textbf{Demonstration:}
Figure \ref{fig:rbo} shows the RBO of the tag distributions of resources over
time for our people tagging dataset. The RBO value between the tag distribution
after $N$ and $N+M$ tag assignments is high if the $M$ new tag assignments do
not change the ranking of the (top-weighted) tags.
One can see from Figure~\ref{fig:rbo} that the RBO of a randomly generated tag
distribution is pretty low and increases slowly as more and more tags are added over
time.
On the contrary, the RBO of  real tag distributions increases as more and more tags
are added. At the beginning, it increases quickly and remains relative stable
after few thousand tag assignments.
This indicates that the RBO measure allows identifying a consensus in the tag
distributions which may emerge over time and which \emph{goes beyond what one
would expect from a random tagging process}.
A random tagging process produces relative tag proportions which are all very
similar (i.e., all tags are equally important or unimportant). Therefore, the probability that the ranking changes after new tag assignments is higher than it is for real tagging streams where users have produced a clear ranking of tags where some tags are much more important than others. Over time, the gap between real tagging streams and random tagging streams will decrease. However, one can see that within the time-window in which real tagging streams semantically stabilize (i.e., few thousand tag assignments) tag distributions produced by a random process are significantly less stable.
Again, we can see that the tag distributions of heavily tagged resources are slightly more stable than those of moderately tagged resources.

In our work, we empirically chose $p=0.9$ which means that the first 10 ranks
have 86\% of the weight of the evaluation. We got similar results when choosing
higher values of $p$.
For example, when choosing $p = 0.98$ the first 50 items get 86\% of the weight.
If one would chose a lower value for p such as $p=0.1$ (or  $p=0.5$) the first
two element would get 99.6\% (or 88.6\%) of the weight. That means, all elements
with a rank lower than two would be almost ignored and therefore the RBO values show more fluctuation.
However, in all our experiments with different $p$ values the RBO of real tag
distributions was significantly higher than the RBO of random tag
distributions.

\textbf{Limitations and Potentials:}
One advantage of RBO is that it handles mutually non-conjoint lists of tags, weights
highly ranked tags more heavily than lower ranked tags, and is monotonic with
increasing depth of evaluation.
A potential limitation of RBO is that it requires to pick the parameter p which
defines the decline in weights - i.e., how top-weighted the RBO measure is.
Which level of top-weightness is appropriate for the tag distributions in
different tagging systems might be a controversial question.
However, our experiments revealed that as long as the parameter $p$ was not
chosen to be small (i.e., $p < 0.5$), the results remained essentially the same.

  \begin{figure}[t!]
 \centering
 \begin{tabular}{c}
\includegraphics[width=0.72\columnwidth]{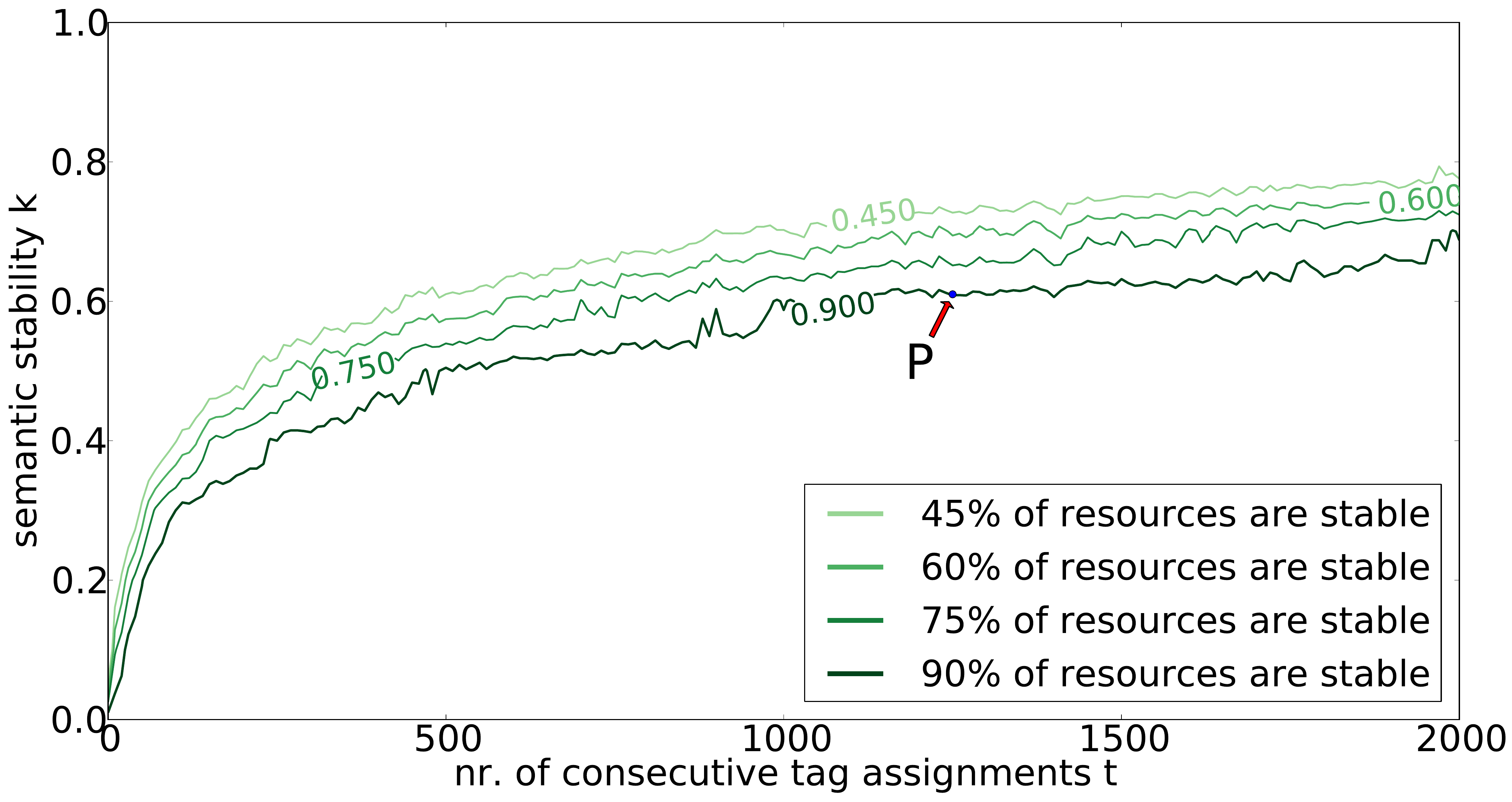}
\end{tabular}
\vspace{-10pt}
\caption{ The percentage of resources (in this case heavily tagged Twitter users) stabilized
at time $t$ with stability threshold $k$. For example, point P indicates that after 1250 tag assignments 90\% of resources exhibit semantic stability (an RBO value) of $0.61$ or higher. }
  \vspace{-10pt}
  \label{fig:comp_contour_twitter_single}
\end{figure}

  \begin{figure*}[ht!]
 \centering
 \begin{tabular}{c}
\includegraphics[width=0.72\textwidth]{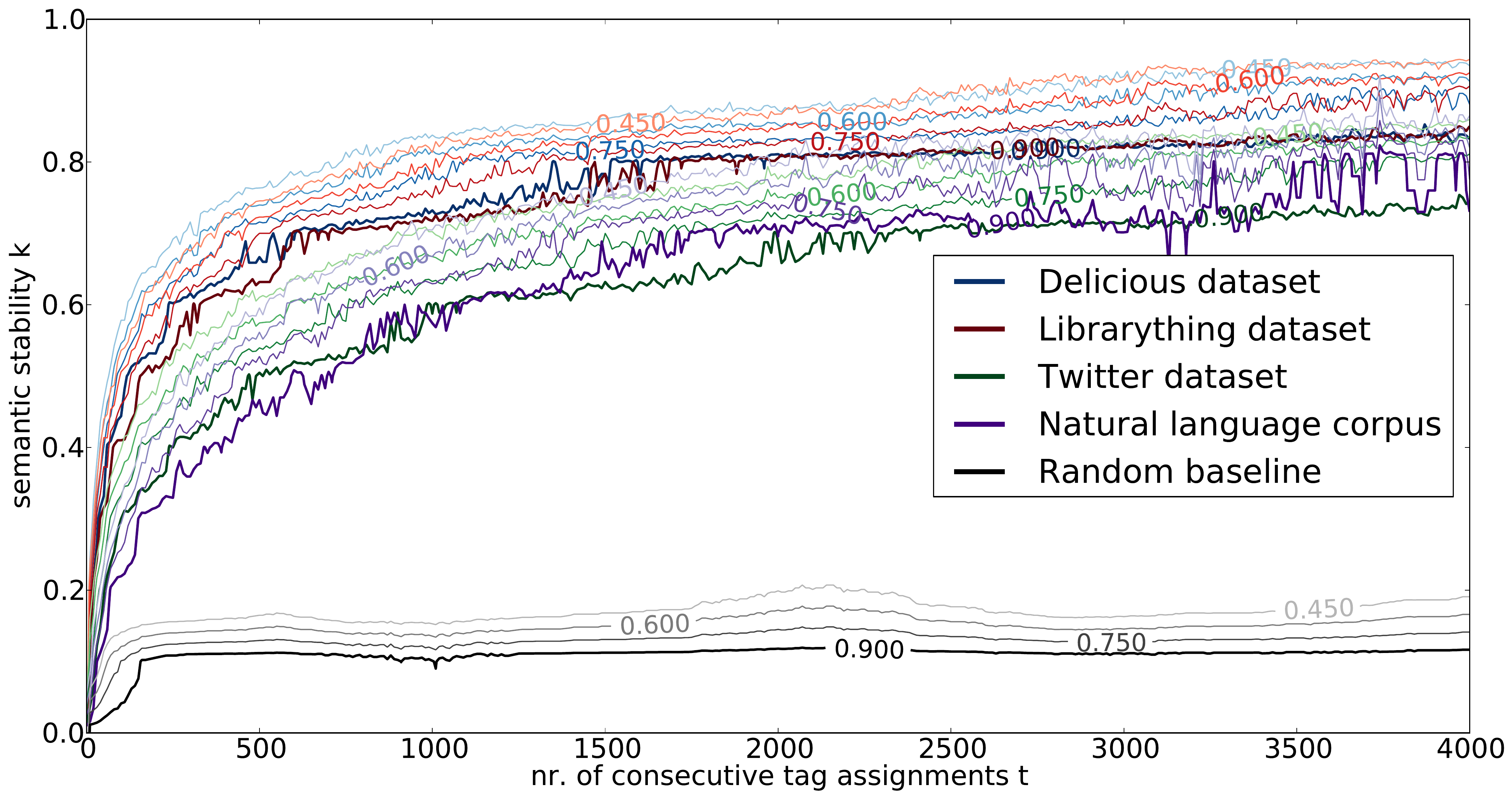}
\end{tabular}
\vspace{-10pt}
\caption{Semantic stabilization of different
social tagging datasets, a natural language corpus and a synthetic random tagging dataset as a control.
The x axis represents the consecutive tag assignments $t$ while the y-axis depicts the RBO (with $p=0.9$) threshold $k$.
The contour lines illustrate the curve for which the function $f(t,k)$ has
constant values. These values are depicted in the lines and represent the
percentage of stabilization $f$. 
 On can see that tagging streams in Delicious and LibraryThing stabilize faster and reach higher levels of semantic stability than other datasets.}
  \vspace{-10pt}
  \label{fig:comp_contour}
\end{figure*}

\subsection{Semantic Stability of Social Tagging\\ Systems}
\label{sec:macro}

Based on the previously defined \emph{Rank Biased Overlap} we propose a method
which allows to investigate the semantic stabilization process in a social tagging system
based on the stabilization process of the individual resources which are tagged. Furthermore, this method
also allows to
compare the extent to which different systems
have become stable.
Given a sample of tagged resources (the sample size $N$ and the type of resources
can be chosen arbitrarily) the goal is to specify how many resources of the
sample have stabilized after a certain number of consecutive tag assignments.
We propose a flexible and fluid definition of the concept of
\emph{stabilization} by introducing (a) a parameter $k$ that constitutes a
threshold for the RBO value and (b) a parameter $t$ that specifies
the number of consecutive tag assignments. We call a resource in a social tagging
system \emph{semantically stable at point $t$}, if the RBO value between its tag
distribution at point $t-1$ and $t$ is equal or greater than $k$.
Our proposed method allows to calculate the percentage of resources that have
semantically stabilized after a number of consecutive tag assignments $t$
according to some threshold for stabilization $k$.
We can define this function by:
\begin{equation}
  f(t,k)=\frac{1}{N}\sum_{i=1}^{N}
  \begin{cases}
    1, & \text{if $RBO(\sigma_{i_{t-1}}, \sigma_{i_t}, p)>k$}.\\
    0, & \text{otherwise}.
  \end{cases}
\end{equation}

We illustrate the semantic stabilization for our sample of heavily tagged
Twitter users in Figure~\ref{fig:comp_contour_twitter_single}. The contour plot
depicts the percentage of resources (i.e., Twitter users) which have become
semantically stable according to some RBO threshold $k$ after $t$ tag
assignments. The figure shows that after 1k tag assignments 90\% of Twitter
users have an RBO value above $0.5$ which can be considered as a medium level
of stability. We define RBO values below $0.4$ as a sign for no stability,
values between $0.4$ and $0.7$ as medium stability and values above $0.7$ as high
stability.

  \begin{figure*}[ht!]
 \centering
 \begin{tabular}{c}
\includegraphics[width=0.72\textwidth]{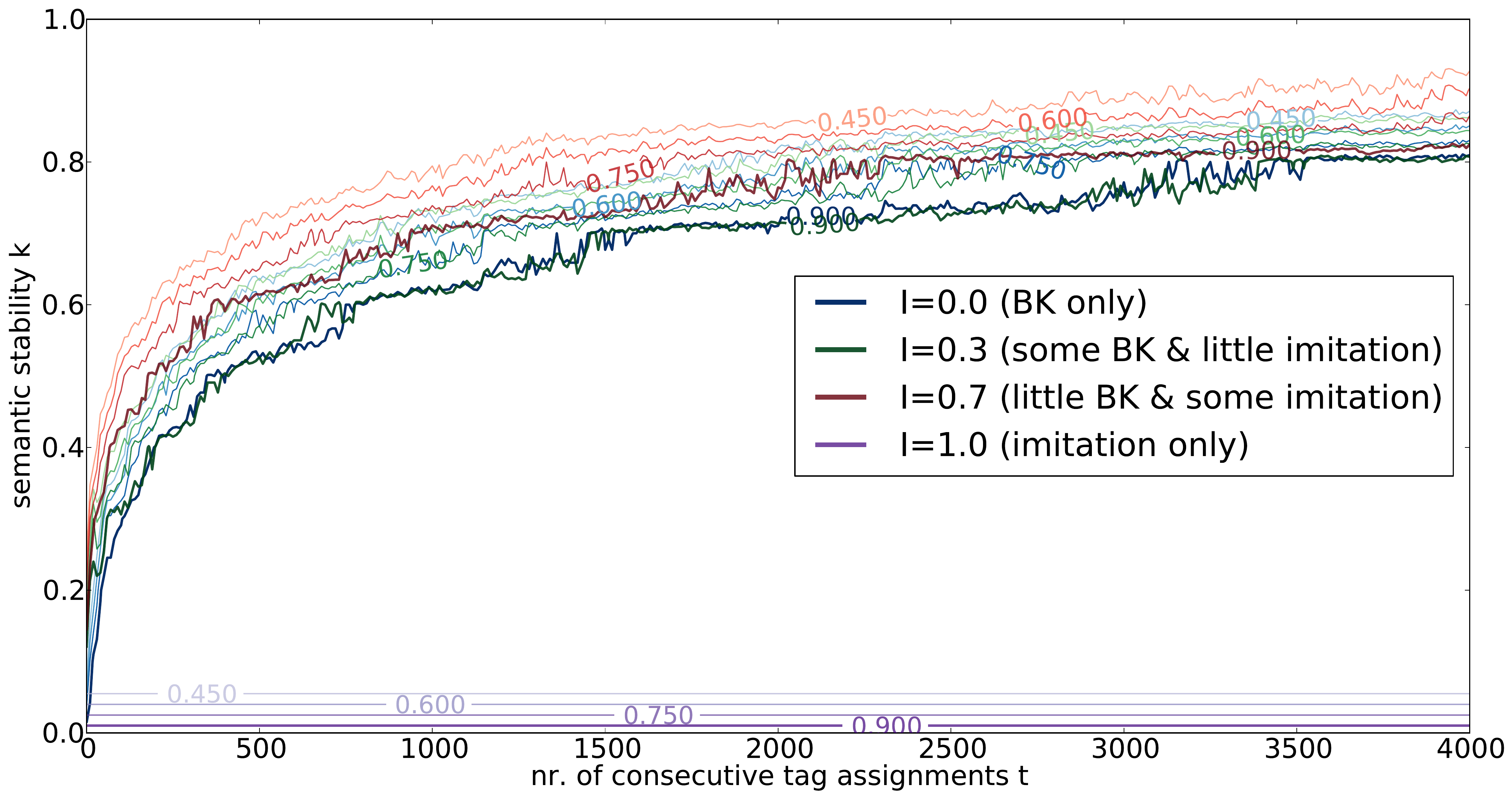}
\end{tabular}
\vspace{-10pt}
\caption{Semantic stabilization of synthetic (i.e. simulated) tagging processes. Tagging streams which are generated by a combination of imitation dynamics (70\%) and background knowledge (30\%) tend to stabilize faster and reach higher levels of stability than streams which are generated by imitation behavior (I=1) or background knowledge (I=0) alone.}
  \label{fig:contour_synthetic}
\vspace{-10pt}
\end{figure*}

\subsubsection{Results \& Discussion}
\label{subsubsec:results}

In this section we use our novel method to compare the semantic stabilization process of different social
tagging systems introduced in Section~\ref{sec:dataset}.

The contour plot in Figure~\ref{fig:comp_contour} depicts the percentage of
resources which have become semantically stable according to some RBO threshold
$k$ after $t$ tag assignments in different social tagging systems.
First of all, we can see that the random dataset exhibits by far the lowest
stabilization since the resources just stabilize for low $k$ ($k < 0.2$) even
after a large amount of tag assignments $t$.
That means, the $k$ threshold for which 90\%, 75\%, 60\% and 45\% of all
resources have an equal or higher $RBO$ values than $k$ is very low.
Contrary, we can see that real-world tagging systems exhibit much higher
stability. The highest (i.e., high $k$ values) and fastest (i.e., low $t$
values) overall tag stabilization can be observed for Delicious and
LibraryThing which both encourage imitation behavior by
suggesting previously assigned tags (see Delicious) and by making previously
assigned tags visible during the tagging process (see LibraryThing).

In Twitter users first have to create a tag (aka user list) and afterwards
select the resources (aka users) to which they want to assign the tag.
During this tagging process, tags which have been previously assigned to users
are not visible and therefore it is unlikely that imitation behavior plays a
major role in Twitter\footnote{If users want to see which other tags have
previously been assigned to a user they need to visit her profile page and
navigate to the list membership section.
Since this is fairly time intensive one can speculate that it is unlikely that
users imitate the previously assigned tags but create their own tags and assign
users to them based on what they know about them and how they want to organize
them.}.
Interestingly, our results show that despite the difference in the user
interfaces, the people tagging streams in Twitter exhibit similar stabilization
patterns as the book and website tagging streams in Delicious and LibraryThing.
However, people tagging streams in Twitter stabilize slightly slower and less
heavily than the tagging streams in Delicious and LibraryThing where imitation
behavior is encouraged.
This result is striking since it suggests that \emph{imitation cannot be the
only factor which causes the stable patterns which arise when a large group of
users tag a resource}.
Our empirical results from different social tagging systems are in line the results from the user study presented in \cite{Bollen2009} which also shows that tag distributions of resources become
stable regardless of the visibility of previously assigned tags. The presence of tag suggestions may provoke
a higher and faster agreement between users who tag a resource and may therefore lead to higher levels of stability, but it is clearly not the
only factor causing stability.
Our results suggest that in tagging system which encourage imitation less than 1k tag assignments are necessary
before a tagging stream becomes semantically stable (i.e., the rank agreement has reached a certain level and does not change anymore), while in tagging systems which do not encourage imitation more than 1k tag assignments are required.
\section{Explaining Semantic Stability}
\label{sec:exsemstab}
The experimental results reported in \cite{Bollen2009} as well as our own empirical results on the people tagging dataset from Twitter suggest that stable
patterns may also arise in the absence of imitation behavior.
As a consequence, other factors that might explain semantic stabilization, such as shared background
knowledge and stable properties of natural language, deserve further
investigation.

\subsection{Imitation and Background Knowledge}
\label{subsec:imitationBK}
To explore the potential impact of imitation and shared background knowledge we simulate the tag choice process.
According to \cite{Dellschaft2008} there are several plausible ways how the tag choice process can be modeled:

\noindent
\textbf{Random tag choice:} Each tag is chosen with the same probability. This
corresponds to users who randomly choose tags from the set of all available tags which seems to be only a plausible strategy for spammers

\noindent
\textbf{Imitation:} The tags are chosen with a probability that is proportional
to the tag's occurrence probability in the previous stream. This selection strategy corresponds
to the Polya Urn model described in \cite{Golder2006} where only tags that have been used
before are in the urn and can be selected.
This corresponds to users who are easily influenced by other users.

\noindent
\textbf{Background Knowledge:} The tags are chosen with a probability that is proportional
to the tag's probability in the shared background knowledge of users. This
corresponds to users who choose tags that seem appropriate based on
their own background knowledge.

In our simulation, we assume that the tag choice of users might be driven by
both imitation and background knowledge.
Similar to the epistemic model \cite{Dellschaft2008}
we introduce a parameter $I$ describing the impact of imitation.
Consequently, the impact of shared background knowledge is $1-I$.
We run $I$ from 0 to 1 -- i.e., we simulate tagging streams which
have been generated by users who only use the imitation strategy to choose their
tags ($I=1$), users who only rely on their background knowledge when
selecting tags ($I=0$), and users who adapt both strategies.
We use a word-frequency
corpus\footnote{\url{http://www.monlp.com/2012/04/16/calculating-word-and-n-gram-statistics-from-a-wikipedia-corpora/}}
from Wikipedia to simulate the shared background knowledge.
For each synthetic dataset we simulate 100 tagging streams in order to have the
same sample size as for our real-world datasets introduced in
Section~\ref{sec:dataset}.

Our results in Figure~\ref{fig:contour_synthetic} show the percentage of
resources which have a RBO value equal or higher than $k$ after $t$ tag
assignments for different synthetic tagging datasets.
One can see from this figure that a synthetic tagging dataset with $I=1$ (i.e.,
a datasets which was solely created via imitation behavior) does not
stabilize over time since more than 90\% of the resources have very low RBO
values (i.e., $k< 0.1$)  also after a few thousand tag assignments. 
This is consistent with our intuition since a model which is purely based on
imitation dynamics fails to introduce new tags and therefore no ranked lists of tags per resource can
be created.

Further, one can see that a synthetic tagging dataset with $I=0$ (i.e., a
tagging datasets which was solely created via background knowledge and therefore
reflects the properties of a natural language system) stabilizes slightly slower
than a synthetic tagging dataset which was generated by a mixture of background
knowledge and imitation dynamics ($I=0.7$).
This is particularly interesting since it suggests that \emph{when shared background knowledge
(encoded in natural language) is combined with social imitation,
tagging streams reach higher levels of semantic stability ($0.7<k<0.8$) quicker (for lower $t$) than if users either only rely on imitation behavior or on background knowledge}.
Our findings are in line with previous research \cite{Dellschaft2008} which
showed that an imitation rate between 60\% and 90\% is best for simulating real
tag streams of resources.
However, as described in Section~\ref{sec:relWork} their work has certain
limitations which we address by (i) exploring a range of different social
tagging systems including one where no tags are suggested and previously
assigned tags are not visible during the tagging process and (ii) studying the
semantic stabilization process over time rather than the shape of the
rank-ordered tag frequency distribution at a single time point.

\subsection{Stability of Natural Language}
\label{subsec:naturallan}
Since tagging systems are natural
language systems, the regularities and the stability of natural language (see e.g., \cite{Zipf1949} and
\cite{cancho01thesmall}) may cause the stable patterns which we observe in
tagging systems.
That means, one can argue that tagging systems become stable because they are
built on top of natural language which itself is stable.

Our results presented in Figure \ref{fig:comp_contour} show that a natural
language corpus (see Section~\ref{sec:dataset}) -- where users talk about a set
of sample resources -- also becomes semantically stable over time and reaches a medium level of stability
(with $k>0.6$ if $t>1,000$).
Also, our simulation results in Figure \ref{fig:contour_synthetic} show that a
synthetic dataset which is only generated via background knowledge ($I=0.0$) and
therefore reflecting the properties of the natural language, becomes semantically stable
over time and reaches a medium level of stability (with $k>0.6$ if $t>1,000$).
In both cases one can see that the stabilization process of natural language
systems clearly differs from the stabilization process of real tagging streams
which are produced in systems supporting imitation and synthetic tagging
streams which are generated by included imitation mechanisms.
The RBO curve of natural language systems is flatter at the beginning than the
RBO curve of tagging streams which are partly generated via imitation mechanisms
which suggests that more word assignments are needed until a high percentage of
resources have RBO values at or above a certain threshold $k$.
The only tagging stream dataset which shows a similar stabilization process as
the natural language dataset is the people tagging dataset obtained from Twitter
which does not support any imitation mechanisms. This suggest, that the
stability of natural language systems can indeed explain a large proportion of
the stability which can be observed in tagging systems where the tagging process
is not really social (i.e., each user annotates a resource separately
without seeing the tags others used) and no imitation dynamics are supported.
However, tagging systems which support the social aspect of tagging by e.g.,
showing tags which have been previously applied by others, exhibit a faster and
higher level of semantic stabilization than tagging systems which do not
implement these social functionalities.
This suggests that the semantic stability which can be observed in \emph{social}
tagging systems goes beyond what one would expect from natural language systems
and that higher and faster degree of stability is achieved through the social
dynamics in tagging systems; concretely, the imitation behavior of users.

%

\section{Discussion}
\label{sec:discuss}

The main implications of our work are:
(i) We highlight limitations of existing methods for measuring semantic
stability in social tagging streams and introduce a new and more robust method.
However, our method is not limited to social tagging systems and tagging streams
and can be used to measure stability and user agreement in other types of data
streams which are collectively created by a set of users (e.g., hashtag-streams
in Twitter or Wikipedia edit-streams).
(ii) Our simulation results suggest that when aiming to improve semantic
stability of social tagging systems, system designers can exploit the insights
gained from our work by implementing mechanisms which - for example - augment
imitation in 70\% of cases (e.g., by suggesting or showing previously assigned
tags) while tapping into the background knowledge of users in 30\% of cases
(e.g., by requiring users to tag without recommendation mechanisms at place,
thereby utilizing background knowledge).

In future we also want to explore the
lowest number of users that need to tag a resource in order to produce a stable
tag description of the resource for which we would also need to model the number
of tags users simultaneously assign to resources into our experiments.
 Further, we want to point out that for the sake of simplicity we
used the same background knowledge corpus for all resources and neglected the impact of the user interface (i.e., the number of suggested tags and the number of
previously used tags from which they are chosen) on the imitation process.
These user interface parameters are different for distinct tagging systems and have been
varied over time.
Without exactly knowing how  the user interface looked like during the tagging
process and how the algorithm for suggesting and displaying tags worked, it is difficult to properly choose these parameters.

\section{Conclusions}
\label{sec:conclusions}

Based on an in-depth analysis of existing methods, we have presented a novel
method for assessing semantic stabilization processes. We have applied our
method to different social tagging systems empirically, and to different
synthetic tagging streams via simulations. Our results reveal that semantic
stability in tagging systems cannot solely be explained by imitation behavior of
users, rather a \emph{combination of} imitation and background knowledge
exhibits highest semantic stabilization. Summarizing, our work makes
contributions on three different levels.

\emph{Methodological}: 
Based on systematic investigations we identify potentials and limitations of
existing methods for asserting semantic stability in social tagging systems.
Using these insights, we present a novel, yet flexible, method which allows to
measure and compare the semantic stabilization of different tagging systems in a robust way. Flexibility is
achieved through the provision of two meaningful parameters, robustness is
demonstrated by applying it to random control processes. Our method is general
enough to be applicable beyond social tagging systems, e.g., to streams of hashtags on Twitter.

\emph{Empirical}: We conduct large-scale empirical
analyses of semantic stabilization in a series of distinct social tagging
systems using our method. We find that semantic stabilization of tags varies
across diverse systems that adopt different tagging mechanics, which requires
deeper explanations of the dynamics of underlying stabilization processes.

\emph{Explanatory}: We investigate factors which may explain stabilization
processes in social tagging systems using simulations. Our results show that
tagging streams which are generated by a \emph{combination of} imitation
dynamics and shared background knowledge exhibit faster and higher semantic
stability than tagging streams which are generated via imitation dynamics or
natural language phenomena alone.

Our findings are relevant for researchers interested in developing more
sophisticated methods for assessing semantic stability of tagging streams and
for practitioners interested in assessing the extent of semantic stabilization
in social tagging systems on a system scale.




\footnotesize
\smallskip 
\noindent
\textbf{Acknowledgments.}
We thank Dr. William Webber for assistance with his RBO metric and Dr. Harry Halpin for assistance with his semantic stability measure.
This work is in part funded by the FWF Austrian Science Fund Grant
I677. Claudia Wagner is a recipient of a DOC-fForte fellowship of the Austrian Academy of Science.

\balance

\newpage

\footnotesize
\bibliographystyle{abbrv}
\bibliography{hp-biblio}

\begin{thebibliography}{10}

\bibitem{Alstott2013}
J.~Alstott, E.~Bullmore, and D.~Plenz.
\newblock powerlaw: a python package for analysis of heavy-tailed
  distributions.
\newblock 2013.

\bibitem{Bollen2009}
D.~Bollen and H.~Halpin.
\newblock The role of tag suggestions in folksonomies.
\newblock In {\em Proceedings of the 20th ACM conference on Hypertext and
  hypermedia}, HT '09, pages 359--360, New York, NY, USA, 2009. ACM.

\bibitem{cattuto2006}
C.~Cattuto, V.~Loreto, and L.~Pietronero.
\newblock Semiotic dynamics in online social communities.
\newblock In {\em In The European Physical Journal C}, pages 33--37.
  Springer-Verlag, 2006.

\bibitem{Cattuto06}
C.~Cattuto, V.~Loreto, and L.~Pietronero.
\newblock Semiotic dynamics in online social communities.
\newblock In {\em In The European Physical Journal C (accepted}, pages 33--37.
  Springer-Verlag, 2006.

\bibitem{cattuto2007}
C.~Cattuto, V.~Loreto, and L.~Pietronero.
\newblock Semiotic dynamics and collaborative tagging.
\newblock {\em Proceedings of the National Academy of Sciences},
  104(5):1461--1464, 2007.

\bibitem{Chomsky63}
N.~Chomsky and G.~Miller.
\newblock {Finitary Models of Language Users}.
\newblock In Luce, Bush, and Galanter, editors, {\em Handbook of Mathematical
  Psychology 2}, volume Handbook of Mathematical Psychology 2, pages 419--491,
  New York, New York, 1963. Wiley and Sons.

\bibitem{Clauset2009}
A.~Clauset, C.~R. Shalizi, and M.~E.~J. Newman.
\newblock Power-law distributions in empirical data.
\newblock {\em SIAM Rev.}, 51(4):661--703, Nov. 2009.

\bibitem{Cohen1997}
A.~Cohen, R.~N. Mantegna, and S.~Havlin.
\newblock Numerical analysis of word frequencies in artificial and natural
  language texts.
\newblock {\em Fractals}, 1997.

\bibitem{Dellschaft2008}
K.~Dellschaft and S.~Staab.
\newblock An epistemic dynamic model for tagging systems.
\newblock In {\em HT '08: Proceedings of the nineteenth ACM conference on
  Hypertext and hypermedia}, pages 71--80, New York, NY, USA, 2008. ACM.

\bibitem{cancho01thesmall}
R.~Ferrer, Cancho, and R.~V. Sol{\'e}.
\newblock The small world of human language.
\newblock {\em Proceedings of The Royal Society of London. Series B, Biological
  Sciences}, 268:2261--2266, 2001.

\bibitem{Cancho2010}
R.~Ferrer-i Cancho and B.~Elvev\r{a}g.
\newblock {Random Texts Do Not Exhibit the Real Zipf's Law-Like Rank
  Distribution}.
\newblock {\em PLoS ONE}, 5(3):e9411+, Mar. 2010.

\bibitem{Fu2010}
W.-T. Fu, T.~Kannampallil, R.~Kang, and J.~He.
\newblock Semantic imitation in social tagging.
\newblock {\em ACM Trans. Comput.-Hum. Interact.}, 17(3):12:1--12:37, July
  2010.

\bibitem{Golder2006}
S.~Golder and B.~A. Huberman.
\newblock Usage patterns of collaborative tagging systems.
\newblock {\em Journal of Information Science}, 32(2):198--208, April 2006.

\bibitem{Gorlitz2008}
O.~G\"{o}rlitz, S.~Sizov, and S.~Staab.
\newblock Pints: peer-to-peer infrastructure for tagging systems.
\newblock In {\em Proceedings of the 7th international conference on
  Peer-to-peer systems}, IPTPS'08, pages 19--19, Berkeley, CA, USA, 2008.
  USENIX Association.

\bibitem{Gruber1995}
T.~R. Gruber.
\newblock Toward principles for the design of ontologies used for knowledge
  sharing.
\newblock {\em Int. J. Hum.-Comput. Stud.}, 43(5-6):907--928, Dec. 1995.

\bibitem{Halpin2007}
H.~Halpin, V.~Robu, and H.~Shepherd.
\newblock The complex dynamics of collaborative tagging.
\newblock In {\em Proceedings of the 16th international conference on World
  Wide Web}, WWW '07, pages 211--220, New York, NY, USA, 2007. ACM.

\bibitem{Kello2010}
C.~T. Kello, G.~D.~A. Brown, R.~Ferrer-i Cancho, J.~G. Holden,
  K.~Linkenkaer-Hansen, T.~Rhodes, and G.~C. Van~Orden.
\newblock {Scaling laws in cognitive sciences}.
\newblock {\em Trends in Cognitive Sciences}, 14(5):223--232, May 2010.

\bibitem{Kipp2006}
M.~E.~I. Kipp and G.~D. Campbell.
\newblock Patterns and inconsistencies in collaborative tagging systems : An
  examination of tagging practices.
\newblock Nov. 2006.

\bibitem{Korner2010}
C.~K\"{o}rner, D.~Benz, A.~Hotho, M.~Strohmaier, and G.~Stumme.
\newblock Stop thinking, start tagging: tag semantics emerge from collaborative
  verbosity.
\newblock In {\em Proceedings of the 19th international conference on World
  wide web}, WWW '10, pages 521--530, New York, NY, USA, 2010. ACM.

\bibitem{Li92randomtexts}
W.~Li.
\newblock Random texts exhibit zipf's-law-like word frequency distribution.
\newblock {\em IEEE Transactions on Information Theory}, pages 1842--1845,
  1992.

\bibitem{Lin2012}
N.~Lin, D.~Li, Y.~Ding, B.~He, Z.~Qin, J.~Tang, J.~Li, and T.~Dong.
\newblock The dynamic features of delicious, flickr, and youtube.
\newblock {\em J. Am. Soc. Inf. Sci. Technol.}, 63(1):139--162, Jan. 2012.

\bibitem{Lorince2013}
J.~Lorince and P.~M. Todd.
\newblock Can simple social copying heuristics explain tag popularity in a
  collaborative tagging system?
\newblock In {\em Proceedings of the 5th Annual ACM Web Science Conference},
  WebSci '13, pages 215--224, New York, NY, USA, 2013. ACM.

\bibitem{Macgregor04}
G.~Macgregor and E.~McCulloch.
\newblock Collaborative tagging as a knowledge organisation and resource
  discovery tool.
\newblock {\em Library Review}, 55(5), in press.

\bibitem{Mathes2004}
A.~Mathes.
\newblock Folksonomies: Cooperative classification and communication through
  shared metadata.
\newblock
  \url{http://www.adammathes.com/academic/computer-mediated-communication/folksonomies.html},
  June 2004.
\newblock Accessed: 2013-07-11.

\bibitem{Mika2007}
P.~Mika.
\newblock Ontologies are us: A unified model of social networks and semantics.
\newblock {\em Web Semant.}, 5(1):5--15, Mar. 2007.

\bibitem{montemurro2002}
M.~A. Montemurro and D.~Zanette.
\newblock Frequency-rank distribution of words in large text samples:
  phenomenology and models.
\newblock {\em Glottometrics}, 4:87--99, 2002.

\bibitem{Rapoport1982}
A.~Rapoport.
\newblock {\em Zipf's law revisited}.
\newblock Studienverlag Bockmeyer, 1982.

\bibitem{Schmitz06miningassociation}
C.~Schmitz, A.~Hotho, R.~J{\"a}schke, and G.~Stumme.
\newblock Mining association rules in folksonomies.
\newblock In {\em DATA SCIENCE AND CLASSIFICATION: PROC. OF THE 10TH IFCS
  CONF., STUDIES IN CLASSIFICATION, DATA ANALYSIS, AND KNOWLEDGE ORGANIZATION},
  pages 261--270. Springer, 2006.

\bibitem{sen2006}
S.~Sen, S.~K. Lam, A.~M. Rashid, D.~Cosley, D.~Frankowski, J.~Osterhouse, F.~M.
  Harper, and J.~Riedl.
\newblock tagging, communities, vocabulary, evolution.
\newblock In {\em Proceedings of the 2006 20th anniversary conference on
  Computer supported cooperative work}, CSCW '06, pages 181--190, New York, NY,
  USA, 2006. ACM.

\bibitem{Specia2007}
L.~Specia and E.~Motta.
\newblock Integrating folksonomies with the semantic web.
\newblock In {\em Proceedings of the 4th European conference on The Semantic
  Web: Research and Applications}, ESWC '07, pages 624--639, Berlin,
  Heidelberg, 2007. Springer-Verlag.

\bibitem{Wagner2013HP}
C.~Wagner, S.~Asur, and J.~Hailpern.
\newblock Religious politicians and creative photographers: Automatic user
  categorization in twitter.
\newblock In {\em ASE/IEEE International Conference on Social Computing
  (SocialCom2013)}, 2013.

\bibitem{webber2010}
W.~Webber, A.~Moffat, and J.~Zobel.
\newblock A similarity measure for indefinite rankings.
\newblock {\em ACM Trans. Inf. Syst.}, 28(4):20:1--20:38, Nov. 2010.

\bibitem{Yule1925}
G.~U. Yule.
\newblock {A Mathematical Theory of Evolution, Based on the Conclusions of Dr.
  J. C. Willis, F.R.S.}
\newblock 213(402-410):21--87, Jan. 1925.

\bibitem{Zipf1949}
G.~K. Zipf.
\newblock {\em Human behavior and the principle of least effort}.
\newblock Addison-Wesley Press, 1949.

\bibitem{Zubiaga2011}
A.~Zubiaga, C.~K\"{o}rner, and M.~Strohmaier.
\newblock Tags vs shelves: from social tagging to social classification.
\newblock In {\em Proceedings of the 22nd ACM conference on Hypertext and
  hypermedia}, HT '11, pages 93--102, New York, NY, USA, 2011. ACM.

\end{thebibliography}

\end{document}